\documentclass[a4paper,11pt]{article}
\usepackage{jheppub}
\usepackage[T1]{fontenc}
\usepackage{tikz}
\usetikzlibrary{arrows.meta,decorations.pathreplacing}
\definecolor{torusblue}{RGB}{48,84,125}
\definecolor{torusteal}{RGB}{48,120,113}
\definecolor{torusgold}{RGB}{164,118,53}

\newcommand{\be}{\begin{equation}}
\newcommand{\ee}{\end{equation}}
\newcommand{\bes}{\begin{equation*}}
\newcommand{\ees}{\end{equation*}}
\newcommand{\beq}{\begin{equation}}
\newcommand{\eeq}{\end{equation}}

\newcommand{\bea}{\begin{eqnarray}}
\newcommand{\ea}{\end{eqnarray}}
\newcommand{\barr}{\begin{array}}
\newcommand{\earr}{\end{array}}

\def\d{\mathrm d}
\def\i{{\mathrm i}}

\def\U{{\mathrm{U}}}

\def\GL{{\mathrm{GL}}}
\newcommand{\Tr}{\operatorname{Tr}}

\newcommand{\C}{\mathbb{C}}
\newcommand{\R}{\mathbb{R}}
\newcommand{\norm}[1]{\left\lVert#1\right\rVert}

\title{BPS black holes in flat space from quantum mechanics}
\author{Xiaoyi Shi}
\author{and Gustavo J Turiaci}
\affiliation{Department of Physics, University of Washington, Seattle, WA 98195, USA}
\emailAdd{xiaoys5@uw.edu}
\emailAdd{turiaci@uw.edu}
\abstract{Despite extensive work on BPS black holes in 4d flat space, there is no direct microstate counting from D-brane quantum mechanics in regimes described by Einstein gravity. We derive an exact formula counting BPS configurations of the D-brane system corresponding to black holes in $\mathcal{N}=8$ supergravity obtained from type II strings on $T^6$. As part of the derivation, we relate the BPS equations to a solvable modified Calogero-Moser system. The result agrees with the prediction from the modified elliptic genus of the D1-D5 system, providing a highly nontrivial test of the string dualities relating the two descriptions. We emphasize that this quantum mechanical system is, as a model of quantum black holes, simpler than BFSS and propose some numerical tests to explore this.}

\begin{document}
\maketitle
\flushbottom

\newpage
\section{Introduction}

Recent years have seen remarkable progress in our understanding of supersymmetric black hole microstates. Despite past successes, several foundational questions remained open over the years that are starting to be addressed. We highlight a few examples below.

First, the semiclassical thermodynamic description of black holes has long been known to break down near extremality \cite{Preskill:1991tb}. This poses an apparent obstacle to understanding supersymmetric black holes, which are necessarily extremal. Recent work has addressed this problem by incorporating quantum gravitational fluctuations that become large in this regime and modify the spectrum and dynamics \cite{Iliesiu:2020qvm}. The supersymmetric extension of this analysis connects the resulting gravitational description to microscopic BPS state counting \cite{Heydeman:2020hhw}. Second, microscopic counting typically relies on protected quantities such as the Witten index, raising the question of how to compute the same observables in gravity. Recent work has clarified how to evaluate supersymmetric indices directly from the gravitational path integral \cite{Cabo-Bizet:2018ehj,Iliesiu:2021are}. Third, there was no understanding on how the space of BPS states is distinguished between weak coupling and strong coupling when gravity applies. Recent proposals based on BPS chaos \cite{Lin:2022zxd,Chen:2024oqv} and fortuity \cite{Chang:2024zqi} offer diagnostics to distinguish BPS sectors associated with macroscopic black holes.

Progress has also been made in computing supersymmetric indices for black holes in four-dimensional asymptotically flat space using the gravitational path integral \cite{Boruch:2023gfn,Chen:2024gmc,Boruch:2025biv}, including those arising from type II compactifications on Calabi-Yau threefolds. Despite these advances on the gravity side, a bottleneck is the lack of an explicit microstate counting which requires using the D-brane formulation. Some initial attempts are in \cite{Gaiotto:2005rp, Gaiotto:2006wm, Denef:2007vg} and more recently \cite{Alexandrov:2022pgd, Alexandrov:2023ltz,Alexandrov:2023zjb}. 

The current situation is roughly as follows. For type II compactifications on $T^6$ and $\mathrm{K3}\times T^2$, yielding $\mathcal{N}=8$ and $\mathcal{N}=4$ supergravity respectively, detailed index formulas are available, but their relation to the original D-brane systems relies on string dualities and the $4d/5d$ lift. For example, in the $\mathcal{N}=8$ case, dualities predict that the BPS degeneracies of the open-string D-brane quantum mechanics are encoded in the generating function \cite{Maldacena:1999bp,Shih:2005qf,Pioline:2005vi}
\beq
-\frac{\vartheta_1(z|\tau)^2}{\eta(\tau)^6}
=
\sum_{n\geq0}\sum_{\ell\in\mathbb Z}
\widehat c(4n-\ell^2)\,q^ny^\ell,
\qquad q=e^{2\pi \i\tau},\quad y=e^{2\pi \i z}.
\label{eq:thetas}
\eeq
where $n$ and $\ell$ are combinations of charges in the 4d problem. The relevant elliptic function is given by
$
\vartheta_1(z|\tau)
=-\i\sum_{a\in\mathbb Z}(-1)^a
q^{(a+1/2)^2/2}y^{a+1/2}.
$
Direct computations in the corresponding D-brane quantum mechanics remain limited. The $\mathcal{N}=8$ calculations in \cite{Chowdhury:2014yca,Chowdhury:2015gbk,Chowdhury:2023wss,Chowdhury:2026noq} address low charges, outside the regime of validity of Einstein gravity. Our goal is to prove this formula by directly counting BPS vacua in the quantum mechanics governing open strings between the D-branes. 

Following \cite{Chowdhury:2014yca}, we begin with the simplest example of the quantum mechanics of three D2-branes, each wrapping a different $T^2$ factor of $T^6$, together with $N$ D6-branes wrapping the entire six-torus. The main result of this paper is to explain how to do the counting efficiently and show that the number of BPS vacua is
\begin{eqnarray}
    d_N &=&  \sum_{\substack{r\ge0\\r(r+1)\le N}} \, \,
\sum_{\substack{n_1,\ldots,n_6\geq0\\
n_1+\cdots+n_6=N-r(r+1)}}
2\,p(n_1)\ldots p(n_6). \nonumber\\
&=&- \widehat c(4N).
\end{eqnarray}
In the first line, $p(n)$ are the number of partitions of $n$.  A physical interpretation of the first line is the following. BPS vacua are arranged as a strongly bound state of the D2's and $r(r+1)$ D6-branes, weakly coupled to six stacks of $n_1,\ldots n_6$ D6-branes such that their total number is $N$. The summand counts the number of such configurations. The proof exploits a suitable regime of $B$-field background, in which the BPS vacua are isolated and can be counted explicitly. One can easily show that the first line is equal to $-\widehat{c}(4N)$, the Fourier coefficients that appear in \eqref{eq:thetas}. This is a highly nontrivial test of the string dualities involved in this prediction.

This work suggests two main directions for future research. The first is to extend our methods to $\mathcal{N}=4$ compactifications and, more ambitiously, to $\mathcal{N}=2$ compactifications on Calabi-Yau threefolds. For a generic Calabi-Yau, direct microscopic counting remains far less developed than in theories with higher supersymmetry. Systematic methods for counting BPS vacua in the corresponding D-brane quantum mechanics could provide rigorous microscopic results and enable new tests of gravitational predictions. 

A second, potentially more intriguing direction concerns matrix-model holography. The extensively studied BFSS quantum mechanics describes black D0-branes in type IIA in the strongly coupled regime of its 't Hooft limit and is conjectured to describe M-theory in flat space in an appropriate large-$N$ limit. Its flat directions, however, give rise to a continuous spectrum that complicates the study of black-hole thermodynamics. The BMN deformation lifts these directions through supersymmetric mass terms, yielding a discrete spectrum. This comes at the cost of a more complicated gravitational description, with dual black hole geometries that are not known. The model studied here, and a generalization with multiple D2-branes, first constructed explicitly in \cite{Chowdhury:2014yca}, provides another concrete quantum mechanical system with a proposed gravitational dual. It offers two potential advantages. First, the corresponding black holes retain a macroscopic horizon in the low temperature limit. Second, generic deformations of the torus metric and $B$-field lift the continuous BPS moduli, once the universal collective coordinates are removed. These features may make the model more amenable to numerical study. In the conclusions, we propose tests of the duality through unprotected observables, including correlation functions.

The paper is organized as follows. In section~\ref{sec:qm-review}, we review the D-brane quantum mechanics of the D2/D2/D2/D6 system and present the BPS equations in the simplest possible form. Section~\ref{sec:counting} constructs the BPS vacua in the presence of a small NS-NS $B$-field on the six-torus. It first solves the BPS equations in the limit where the parameters induced by the $B$-field vanish, and then shows that a small $B$-field lifts the remaining moduli. In section~\ref{sec:SummaryofCounting}, we show that its generating function agrees with the prediction from string dualities. Finally, section~\ref{sec:conclusion} discusses future directions, including possible numerical tests of the proposed duality with the black hole.

\section{The D-brane quantum mechanics}\label{sec:qm-review}

In this section, we describe the supersymmetric quantum mechanics of open strings on D2 and D6 branes wrapped on cycles of $T^6$. We then reduce its BPS equations to a form suited to the counting analysis done in the rest of the paper. 

\subsection{Low energy degrees of freedom}

\begin{figure}[t]
\centering
\begin{minipage}[c]{0.60\linewidth}
\centering
\begin{tikzpicture}[x=1cm,y=1cm,font=\small]
\foreach \xx/\col/\labeltor/\coord/\brane in
 {0/torusblue/{T_1^2}/{x^4,x^5}/{\mathrm{D2}_1},
  2.55/torusteal/{T_2^2}/{x^6,x^7}/{\mathrm{D2}_2},
  5.1/torusgold/{T_3^2}/{x^8,x^9}/{\mathrm{D2}_3}} {
 \begin{scope}[shift={(\xx,0)}]
  \path[draw=\col,fill=\col!10,line width=.8pt]
   (0,0) ellipse (1.05 and .57);
  \path[draw=\col,fill=white,line width=.8pt]
   (-.48,.02) .. controls (-.27,.29) and (.27,.29) .. (.48,.02)
   .. controls (.27,-.12) and (-.27,-.12) .. (-.48,.02);
  \draw[\col,line width=.6pt]
   (-.48,.02) .. controls (-.27,-.30) and (.27,-.30) .. (.48,.02);
  \node at (0,1.00) {$\labeltor$};
  \node[font=\scriptsize] at (0,-.93) {$\coord$};
  \node[text=\col,font=\scriptsize] at (0,-1.30) {$\brane$};
 \end{scope}
}
\node at (1.275,0) {$\times$};
\node at (3.825,0) {$\times$};
\draw[decorate,decoration={brace,mirror,amplitude=5pt},gray]
 (-1.02,-1.60)--(6.12,-1.60);
\node[font=\small] at (2.55,-2.05) {$N\ \mathrm{D6}:\ T_1^2\times T_2^2\times T_3^2$};
\end{tikzpicture}
\end{minipage}\hfill
\begin{minipage}[c]{0.36\linewidth}
\centering
\begin{tikzpicture}[x=.88cm,y=.88cm,>=Stealth,font=\small,
 brane/.style={circle,draw,line width=.8pt,minimum size=9mm,inner sep=1pt}]
 \node[brane,draw=torusblue,fill=torusblue!8] (n1) at (0,1.7) {$1$};
 \node[brane,draw=torusteal,fill=torusteal!8] (n2) at (-1.72,-1.3) {$2$};
 \node[brane,draw=torusgold,fill=torusgold!8] (n3) at (1.72,-1.3) {$3$};
 \node[brane,fill=black!6,minimum size=13mm,align=center] (n4) at (0,-.30)
  {$4$\\[-2pt]\scriptsize $U(N)$};
 \foreach \a/\b in {n1/n2,n2/n3,n3/n1,n1/n4,n2/n4,n3/n4}{
  \draw[->,gray!85,bend left=7,shorten >=2pt,shorten <=2pt] (\a) to (\b);
  \draw[->,gray!85,bend left=7,shorten >=2pt,shorten <=2pt] (\b) to (\a);
 }
 \node[font=\scriptsize,above=3pt] at (n1.north) {$U(1)$};
 \node[font=\scriptsize,below=3pt] at (n2.south) {$U(1)$};
 \node[font=\scriptsize,below=3pt] at (n3.south) {$U(1)$};
\end{tikzpicture}
\end{minipage}
\caption{Left: the three two-tori and the wrapped branes. Right: the open-string quiver. Each pair of nodes carries two
independent chiral fields, one for each arrow orientation. There are also three
adjoint chiral multiplets at each node.}
\label{fig:geometry-quiver}
\end{figure}
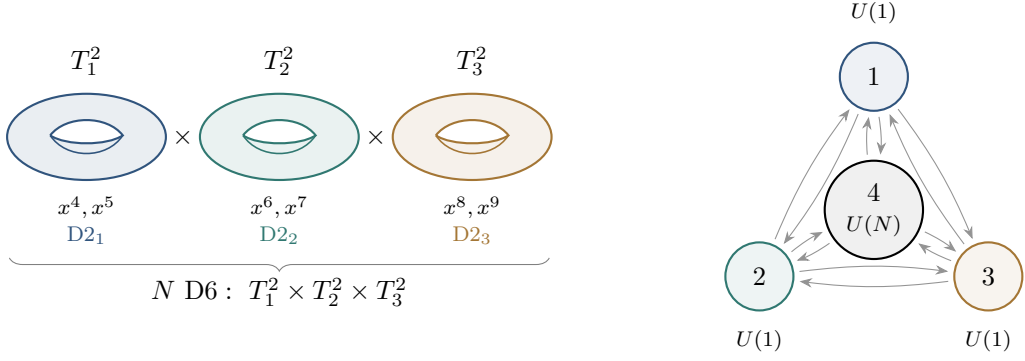

We work in type IIA string theory on $\mathbb{R}^{1,3}\times T^6$, initially taking the torus to be flat with vanishing NS-NS two-form $B$. Following \cite{Chowdhury:2014yca,Chowdhury:2015gbk}, we will later introduce small deformations of the torus metric and a constant background $B$-field. We denote the noncompact coordinates by $x^0,\ldots,x^3$ and the compact coordinates by $x^4,\ldots,x^9$, and organize the latter into three two-tori
\begin{equation}
T^6=T_1^2\times T_2^2\times T_3^2,
\qquad T_1^2=(x^4,x^5),\quad T_2^2=(x^6,x^7),\quad
T_3^2=(x^8,x^9).
\label{eq}
\end{equation}
We consider four stacks of wrapped D-branes, labeled by $k=1,2,3,4$, with $N_k$ branes in each stack, as illustrated in figure~\ref{fig:geometry-quiver}. For $k=1,2,3$, stack $k$ consists of $N_k$ D2-branes wrapping $T_k^2$, while stack $4$ consists of $N_4$ D6-branes wrapping the entire six-torus. We label the configurations by the integers $(N_1,N_2,N_3,N_4)$. Throughout this paper, we specialize to
\beq
N_1=N_2=N_3=1,\qquad N_4=N,
\eeq
or equivalently $(1,1,1,N)$, leaving other choices of brane multiplicities for future work.

The low-energy dynamics of open strings on this D-brane configuration is described by an $\mathcal{N}=4$ quiver quantum mechanics with four real supercharges, whose quiver is shown in figure~\ref{fig:geometry-quiver}. Each stack $k$ supports a $\U(N_k)$ gauge symmetry, giving the total gauge group $\U(1)^3\times\U(N)$. The common diagonal $\U(1)$ acts trivially on all adjoint and bifundamental matter fields, so the effective gauge group is
\beq
G=\frac{\U(1)^3\times\U(N)}{\U(1)_{\mathrm{diag}}}.
\label{eqn:gauge-group}
\eeq
We next describe the field content and its transformation laws under $G$.

\smallskip

The fields of the quiver quantum mechanics are organized into three types of supermultiplets. We describe each through its bosonic components, which is enough to formulate the classical BPS equations. They are:
\begin{enumerate}
\item The vector multiplet at stack $k$ contains three Hermitian $N_k\times N_k$ matrices $X_a^{(k)}$, $a=1,2,3$, describing positions in noncompact space. At the three D2 nodes, $N_k=1$, so these fields reduce to real numbers. At the D6 node we have three $N\times N$ matrices $X_a^{(4)}$.

\item The three adjoint chiral multiplets at stack $k$ contain complex $N_k\times N_k$ scalar matrices $\Phi_i^{(k)}$, $i=1,2,3$. At a D2 node $k\leq3$, $\Phi_k^{(k)}$ combines the two Wilson-line degrees of freedom along the wrapped torus $T_k^2$. For $i\ne k$, $\Phi_i^{(k)}$ instead describes the position of the D2-brane on the transverse torus $T_i^2$. Since the D6-branes wrap the entire six-torus, each $\Phi_i^{(4)}$ combines the two Wilson-line degrees of freedom along $T_i^2$.

\item In the undeformed geometry, each pair of stacks $k$ and $\ell$ has four mixed Neumann-Dirichlet directions. The corresponding open strings give rise to two light chiral multiplets, with complex  $Z^{(k\ell)}$ and $Z^{(\ell k)}$ associated with the two arrow orientations. In our convention, $Z^{(k\ell)}$ maps from node $\ell$ to node $k$:
\begin{equation}
Z^{(k\ell)}: \C^{N_\ell} \longmapsto \C^{N_k}.
\label{eq}
\end{equation}
Thus the six D2-D2 fields are complex numbers, while $Z^{(i4)}$ and $Z^{(4i)}$, $i=1,2,3$, are $1\times N$ rows and $N\times1$ columns, respectively. Fields associated with opposite arrows are independent complex variables, not Hermitian conjugates.
\end{enumerate}

\smallskip 

We now describe how these fields transform under the symmetries of the theory. A gauge transformation specified by $g_k=e^{\mathrm{i}\theta_k}$ for $k=1,2,3$ and $g_4\in\U(N)$ acts as
\begin{equation}
X_a^{(k)} \mapsto g_k X_a^{(k)} g_k^{-1},
\qquad
\Phi_i^{(k)} \mapsto g_k \Phi_i^{(k)} g_k^{-1},
\qquad
Z^{(k\ell)} \mapsto g_k Z^{(k\ell)} g_\ell^{-1}.
\end{equation}
The diagonal subgroup $\U(1)_{\text{diag}}$, defined by $g_1=g_2=g_3=e^{\mathrm{i}\theta}$ and $g_4=e^{\mathrm{i}\theta}I_N$, acts trivially on all these fields. The effective gauge group is therefore the quotient given in \eqref{eqn:gauge-group}. 

The theory also has several global symmetries. Under rotations of the noncompact spatial directions, $X_a^{(k)}$ transforms as a vector, while $\Phi_i^{(k)}$ and $Z^{(k\ell)}$ transform as scalars. The potential is also invariant under six independent complex shifts. For each $i=1,2,3$, these act as
\begin{align}
\Phi_i^{(k)}&\mapsto\Phi_i^{(k)}+\xi_i,
&&k\in\{1,2,3\}\setminus\{i\},\label{eq:position-shifts}\\
\Phi_i^{(i)}&\mapsto\Phi_i^{(i)}+\chi_i,
&\Phi_i^{(4)}&\mapsto\Phi_i^{(4)}+\chi_iI_N,
 \label{eq:Wilson-shifts}
\end{align}
with independent parameters $\xi_i,\chi_i\in\C$ and all remaining fields held fixed. The $\xi_i$ shifts translate the two transverse D2-branes together on $T_i^2$, while the $\chi_i$ shifts change the common Wilson lines of the D2-brane wrapping $T_i^2$ and the D6 stack. The corresponding collective coordinates are compact, reflecting the periodicity of torus positions and Wilson lines. Finally, common translations in noncompact space act as
\begin{equation}
 X_a^{(k)}\longmapsto X_a^{(k)}+a_aI_{N_k},
 \qquad a_a\in\R,\quad a=1,2,3,
 \label{eq:spatial-shifts}
\end{equation}
with the same translation parameter $a_a$ at every node. The degrees of freedom associated with these compact shifts and noncompact translations describe free collective motion of the D-brane system.

\subsection{Interactions}

Having specified the gauge group, matter content, and symmetry transformations, we take the standard kinetic terms for the vector and chiral multiplets. To complete the description of the theory, we now introduce the superpotential and the Fayet-Iliopoulos (FI) parameters. The superpotential determines the F-term potential upon eliminating the auxiliary $F$ fields, while the FI parameters couple linearly to the auxiliary $D$ fields in the vector multiplets. The full off-shell action is given in Appendix C of \cite{Denef:2002ru} or equations (2.32)-(2.35) of~\cite{Hori:2014tda}. In the next section, we restrict attention to BPS configurations.

\subsection*{The superpotential}
\label{sec:fterms}
Let $\phi_\alpha$ denote the complex scalar components of the chiral multiplets. With canonical kinetic terms, eliminating the auxiliary fields $F_\alpha$ from the bosonic action gives
\begin{equation}
S \supset \sum_\alpha \int \d t\left[
|F_\alpha|^2+
\left(F_\alpha\frac{\partial W}{\partial\phi_\alpha}
+\mathrm{c.c.}\right)
\right]
\quad\Rightarrow\quad
V_F=\sum_\alpha\left|\frac{\partial W}{\partial\phi_\alpha}\right|^2,
\label{eq}
\end{equation}
where $\mathrm{c.c.}$ denotes complex conjugation. This defines the F-term potential $V_F$ and supersymmetry fixes the interactions with the fermions. The quantum mechanics of open strings in our D-brane system has the following superpotential.

Following the normalization of~\cite{Chowdhury:2014yca,Chowdhury:2015gbk}, we write the superpotential as the sum of four terms:
\begin{equation}
W=W_{\Phi ZZ}+W_{ZZZ}+W_{\Phi\Phi\Phi}+W_{\rm bg}.
\label{eq:superpotential}
\end{equation}
The first term couples the bifundamental fields to relative positions and Wilson lines:
\begin{equation}
W_{\Phi ZZ}=\sqrt{2}\sum_{\mathrm{cyc}(i,j,k)}
\left[
(\Phi_k^{(i)}-\Phi_k^{(j)})Z^{(ij)}Z^{(ji)}
+Z^{(i4)}(\Phi_i^{(i)}I_N-\Phi_i^{(4)})Z^{(4i)}
\right],
\label{eq:W-PhiZZ}
\end{equation}
where the sum runs over the three cyclic permutations of $(1,2,3)$. These couplings give masses to D2-D2 strings when the branes are separated on their common transverse torus, and to D2-D6 strings when relative Wilson lines are turned on along the torus wrapped by the D2-brane. The second term couples three bifundamental fields forming an oriented triangle in the quiver:
\begin{equation}
W_{ZZZ}=\sqrt{2}\sum_{(k,\ell,m)\in\mathcal T}
(-1)^{\delta_{k1}\delta_{\ell3}\delta_{m4}}
\Tr\left(Z^{(k\ell)}Z^{(\ell m)}Z^{(mk)}\right),
\label{eq:W-triangles}
\end{equation}
where $\mathcal T$ consists of triples of distinct nodes modulo cyclic rotations, represented with the smallest index first. The relative normalization and sign convention, including the exceptional minus sign for the $(1,3,4)$ triangle, follow~\cite{Chowdhury:2015gbk}. The adjoint fields at the D6 node also have a nonabelian cubic interaction:
\begin{equation}
W_{\Phi\Phi\Phi}=-\sqrt{2}\Tr\left(
\Phi_1^{(4)}[\Phi_2^{(4)},\Phi_3^{(4)}]\right).
\label{eq:W-adjoint}
\end{equation}
This term vanishes for $N=1$, when the D6 gauge group is abelian. Analogous adjoint cubic interactions arise at the D2 nodes when any of the ranks $N_1$, $N_2$ or $N_3$ exceed one.

For the undeformed product torus with vanishing $B$-field, the BPS equations admit continuous families of solutions, making direct counting difficult. Following~\cite{Chowdhury:2014yca}, we introduce small, constant deformations of the torus metric and the NSNS two-form $B$ to lift relative flat directions. Their contribution to the superpotential is encoded in six complex parameters $c^{(k\ell)}=c^{(\ell k)}$, with $1\leq k<\ell\leq4$. In terms of the reference complex coordinates $z_i=x^{2i+2}+\mathrm{i}x^{2i+3}$ on $T_i^2$, the correspondence is schematically
\beq
c^{(ij)} \sim g_{z_i z_j},
\qquad
c^{(i4)} \sim \epsilon_{ijk}B_{\bar z_j\bar z_k},
\eeq
where $i\ne j$ in the first relation, and we suppress convention-dependent phases and normalization constants. These deformations generate the linear superpotential term~\cite{Chowdhury:2014yca,Chowdhury:2015gbk}
\begin{equation}
W_{\rm bg}=\sqrt{2}\sum_{\mathrm{cyc}(i,j,k)}
\left[
c^{(ij)}(\Phi_k^{(i)}-\Phi_k^{(j)})
+c^{(i4)}\Tr\left(
\Phi_i^{(i)}I_N-\Phi_i^{(4)}
\right)
\right].
\label{eq}
\end{equation}
The constant $B$-field also induces FI parameters, which we discuss next.

\subsection*{D term equations}
\label{sec:dterms}

The auxiliary fields $D_k$ in the vector multiplets enter the Lagrangian through
\begin{equation}
S\supset\sum_{k=1}^4
\int \d t\,\Tr\!\left[
\frac12 D_k^2+D_k(\mu_k-\zeta_kI_{N_k})
\right],\quad \Rightarrow \quad  V_D=\frac12\sum_{k=1}^4
 \Tr\!\left(\mu_k-\zeta_kI_{N_k}\right)^2,
\end{equation}
where $\zeta_k$ are real FI parameters and the Hermitian moment maps are
\begin{equation}
 \mu_k=\sum_{\ell\ne k}
 \left(
 Z^{(k\ell)}Z^{(k\ell)\dagger}
 -Z^{(\ell k)\dagger}Z^{(\ell k)}
 \right)
 +\sum_{i=1}^3[\Phi_i^{(k)},\Phi_i^{(k)\dagger}].
 \label{eq:moment-map}
\end{equation}
Each bifundamental contributes positively at its target node and negatively at its source node, while the commutators account for the adjoint fields. 

For a $T^6$ without a background $B$-field, the FI parameters would vanish. According to the analysis of appendix A of \cite{Chowdhury:2015gbk}, the FI parameters $\zeta^k$ are given by linear combinations of $B_{z_i \bar{z}_i}$. The explicit form satisfies the constraint
\begin{equation}
 \sum_{k=1}^4N_k\zeta_k
 =\zeta_1+\zeta_2+\zeta_3+N\zeta_4=0.
 \label{eq:FI-constraint}
\end{equation}
Indeed, summing the traces of \eqref{eq:moment-map} cancels the bifundamental contributions, while every commutator is traceless. This is a general consistency condition on the FI parameters.

\subsection*{Scaling}

Section 5 of \cite{Chowdhury:2023wss} argues that there cannot be any higher order corrections to the superpotential. A subtlelty that could complicate the analysis is the presence of periodic identification of the $\Phi$ fields, some because they represent position on the torus, some because they are Wilson lines of compact gauge groups. Following \cite{Chowdhury:2014yca}, we introduce a small parameter $0<\lambda\ll1$ and scale all background parameters uniformly:
\begin{equation}
 c_{\rm phys}^{(ij)}=\lambda c^{(ij)},\qquad
 c_{\rm phys}^{(i4)}=\lambda c^{(i4)},\qquad
 c_{\rm phys}^{(k)}=\lambda c^{(k)},
 \label{eq:parameter-scaling}
\end{equation}
where $c^{(k)}=\zeta_k$ denotes the real FI parameter at node $k$. We simultaneously rescale the fields as
\begin{equation}
 Z_{\rm phys}=\sqrt{\lambda}\,Z,\qquad
 \Phi_{\rm phys}=\sqrt{\lambda}\,\Phi,\qquad
 X_{\rm phys}=\sqrt{\lambda}\,X.
 \label{eq:field-scaling}
\end{equation}
It is easy to see that for the terms we presented so far, both $V_F$ and $V_D$ scale homogeneously as $\lambda^2$. The scaling of $X$ is required for the interactions that appear in the kinetic term. The terms without time derivatives lead to the following potential
\begin{align}
 V_{\rm gauge}={}&
 \sum_{k\ne\ell}\sum_{a=1}^3
 \norm{X_a^{(k)}Z^{(k\ell)}-Z^{(k\ell)}X_a^{(\ell)}}^2
 +\sum_{k=1}^4\sum_{a,i=1}^3
 \norm{[X_a^{(k)},\Phi_i^{(k)}]}^2\notag\\
 &+\frac14\sum_{k=1}^4\sum_{a,b=1}^3
 \norm{[X_a^{(k)},X_b^{(k)}]}^2,
 \label{eq:spatial-potential}
\end{align}
where $\norm{M}^2=\Tr(M^\dagger M)$. Each term is quartic in the fields and therefore scales as $\lambda^2$. The complete leading bosonic potential thus obeys $V \to \lambda^2 V$. As we take $\lambda \to 0 $ we can ignore any periodic identification of the fields. From now on, we work with the rescaled variables.

\subsection{Space of BPS vacua}
Finally, in the rest of the section we focus on the equations specifying BPS configurations and put them in the simplest form possible. 

\subsection*{BPS equations}

The possible vacua of the model has vanishing fermionic fields and therefore can be determined through the bosonic potential. The total potential is given by
\begin{equation}
 V=V_F+V_D+V_{\text{gauge}}.
 \label{eq:total-potential}
\end{equation}
BPS configurations are such that the total potential vanishes. Each individual term in $V_F$, $V_D$ and $V_{\text{gauge}}$ are sums of square and therefore the vacuum is BPS if each individual term vanishes. The F and D contributions to the potential lead to the following algebraic equations for the bosonic matter fields
\begin{equation}
\d W = 0,\qquad \text{and}\qquad \mu_k = \zeta_k I_{N_k},
\end{equation}
where we think of $W$ as a holomorphic function of $\Phi_j^{(k)}$ and $Z^{(ij)}$. The equations arising from the D-term is non-holomorphic. Setting $V_{\text{gauge}}=0$ fixes the vector multiplets
\beq
X_a^{(k)} Z^{(k\ell)} = Z^{(k\ell)} X_a^{(\ell)} ,\qquad [X_a^{(k)},\Phi_i^{(k)}]=0,\qquad [X_a^{(k)},X_b^{(k)}]=0.
\eeq
We are interested in identifying the solutions to these equations, modulo gauge transformations $G$ and modulo the shift symmetries described earlier. 

\subsection*{When vector multiplets decouple}

Our first step to simplify the equations is to show that if $c^{(i4)}\neq 0$, then any BPS configuration can be made to satisfy $X_j^{(k)}=0$ using the symmetries of the theory. To prove this, fix a spatial index $a$ and define the Hermitian matrices
$
 K_i=X_a^{(4)}-X_a^{(i)}I_N,
$
for $i=1,2,3$. The equations from $V_{\text{gauge}}=0$ can be put in the following form
\begin{equation}
 K_iZ^{(4i)}=0,
 \qquad
 [K_i,\Phi_j^{(4)}]=0
 \quad\text{for all }j.
\end{equation}
On the other hand, varying $W$ with respect to $\Phi_i^{(4)}$
gives
\begin{equation}
 Z^{(4i)}Z^{(i4)}
 +[\Phi_j^{(4)},\Phi_k^{(4)}]
 +c^{(i4)}I_N=0,
 \qquad (i,j,k)\text{ cyclic}.
\end{equation}
Multiplying by $K_i^2$ and taking the trace eliminates the
first two terms:
\begin{equation}
 \begin{aligned}
 \Tr\!\left(K_i^2Z^{(4i)}Z^{(i4)}\right)
 &=Z^{(i4)}K_i^2Z^{(4i)}=0,\\
 \Tr\!\left(K_i^2[\Phi_j^{(4)},\Phi_k^{(4)}]\right)
 &=\Tr\!\left([K_i^2,\Phi_j^{(4)}]\Phi_k^{(4)}\right)=0.
 \end{aligned}
\end{equation}
Hence $c^{(i4)}\Tr K_i^2=0$. Since $c^{(i4)}\neq0$ and
$K_i$ is Hermitian, this implies that the matrices themselves have to vanish $K_i=0$. Applying this to all three D2-branes and all spatial indices gives
\begin{equation}
 X_a^{(1)}=X_a^{(2)}=X_a^{(3)}=x_a,
 \qquad
 X_a^{(4)}=x_aI_N.
\end{equation}
We can now use the freedom under rigid translations of the D-brane system in the noncompact space to set $x_a=0$ and therefore $X_j^{(k)}=0$. The nonvanishing of all three $c^{(i4)}$ is a sufficient
condition for this conclusion, not a necessary one.

\smallskip

It is instructive to count the number of equations and unknowns. The fields $\Phi$ and $Z$ contain $3N^2+9$ and $6N+6$ complex components, respectively. Removing six complex shifts and $N^2+2$ real gauge directions leaves $5N^2+12N+16$ real unknowns. Let us now count independent equations. Among  the F equations, holomorphy extends gauge invariance of $W$ to the complexified gauge group, yielding $N^2+2$ complex identities. Together with the six shift identities, these leave $4N^2+12N+14$ potentially independent real F equations. Adding the $N^2+2$ independent real D equations matches the number of unknowns. The naive count therefore predicts $\dim\mathcal M_{\mathrm{BPS}}=0$. We will verify below that this reduced moduli space consists of isolated points.

\smallskip

The dimension count is unchanged if we omit the D-term equations and quotient degrees of freedom by the complexified gauge group. Its additional $N^2+2$ real gauge directions compensate for the redundancies in the superpotential. This motivates the following result.

\subsection*{Ignoring D-terms and complexifying the gauge group}

For $c^{(i4)}\neq0$, $i=1,2,3$, the BPS moduli space is homeomorphic to the space of F-term solutions modulo the complexified gauge group. More precisely, each complex gauge orbit contains a unique unitary gauge orbit satisfying the D-term equations. We now explain this correspondence and verify the conditions required for it to hold in our model. This was heavily used in \cite{Chowdhury:2015gbk,Chowdhury:2023wss,Chowdhury:2026noq}.

Since $W$ is holomorphic, its invariance under $G$ extends to the complexified gauge group
\begin{equation}
G_{\C}
=\frac{(\C^\ast)^3\times\mathrm{GL}(N,\C)}
{\C^\ast_{\mathrm{diag}}},
\end{equation}
where $\C^\ast$ is the multiplicative group of nonzero complex numbers. The group $G_\C$ acts on the fields by the same formulas as $G$. Besides the unitary transformations $e^{\i tH_k}$, complexification introduces transformations of the form
\begin{equation}
g_k(t)=e^{tH_k},
\qquad H_k^\dagger=H_k,
\qquad t\in\R.
\end{equation}
These matrices are Hermitian and positive definite. Their action preserves the F-term equations but generally changes the moment maps $\mu_k$, and hence need not preserve the D-term equations.

Following \cite{Chowdhury:2015gbk,Chowdhury:2023wss,Chowdhury:2026noq}, our counting strategy relies on the homeomorphism\footnote{As stated, this is true for configurations with $\text{gcd}(N_1,N_2,N_3,N_4) =1$ which is certainly our case. For other cases a similar homeomorphism still holds but the quotient on the RHS is more subtle. It is interesting to note that similar arithmetic conditions appear also in the dual D1/D5/KK system for other reasons \cite{Sen:2009gy}.}
\begin{equation}
\mathcal{M}_{\text{BPS}}
=\frac{\{\d W=0\}\cap\{\mu_k=\zeta_kI_{N_k}\}}{G}
\cong
\frac{\{\d W=0\}}{G_{\C}},
\label{eq:complex-gauge-quotient}
\end{equation}
with the same shift directions removed on both sides. This is an application of the Kempf-Ness theorem \cite{King94} and the general topological statement appears in Theorem~4.2(v) of~\cite{Hoskins2014}. Instead of reproducing the full proof, we will just show that $c^{(i4)}\neq0$ for $i=1,2,3$ ensures that within each $G_\C$ orbit of a solution of the F-term equations there is one and only one representative that solves the D-term equations. According to the general theorem, the correspondence preserves the Euler characteristic and, when the solutions are isolated, their number.

The argument is the following. Let $\phi^\alpha$ denote the chiral fields, and let $\norm{\phi}^2$ be the sum of their squared matrix norms. For a fixed F-term solution $\phi$, define the real function
\begin{equation}
\mathcal K_\phi(g)
=\norm{g\cdot \phi}^2
-2\sum_{k=1}^4\zeta_k\log|\det g_k|.
\label{eq:balancing-functional}
\end{equation}
This function is invariant under left multiplication by unitary gauge transformations. The constraint $\sum_kN_k\zeta_k=0$ also ensures invariance under the common rescaling $g_k\mapsto s g_k$, with $s\in\C^\ast$, so $\mathcal K_\phi$ is well defined on $G_\C$. The reason to introduce this function is that its variation
along a complex gauge transformation $g_k(t)=e^{tH_k}g_k$ is
\begin{equation}
 \left.\frac{\d}{\d t}\mathcal K_\phi(g(t))\right|_{t=0}
 =
 2\sum_{k=1}^4
 \Tr\!\left[
 H_k\bigl(\mu_k(g\cdot \phi)-\zeta_kI_{N_k}\bigr)
 \right].
 \label{eq:balancing-variation}
\end{equation}
As a function of the complex gauge transformation, the saddle-points of this function are precisely the solutions of the D-term equations. Finding a D-flat representative
amounts to finding an extremum of~\eqref{eq:balancing-functional}
along a complex gauge orbit.

In a basis where each $H_k$ is diagonal, every field component scales exponentially with $t$. Thus $\mathcal K_\phi(t)$ is a sum of exponentials with nonnegative coefficients and a linear FI term. It is therefore convex and any critical point is unique, up to real gauge transformations, and a minimum, and is the corresponding D-flat representative. To guarantee the existence of a minimum, it suffices to show that $\mathcal K_\phi(t)\to+\infty$ as $t\to\pm\infty$ along every nontrivial Hermitian gauge direction. Establishing this growth requires the F-term equations $\d W(\phi)=0$ satisfied by the initial configuration.

The idea is the following. For concreteness let us focus on $t\to\infty$. When evaluating $\norm{g \cdot \phi}^2$, some components of $\phi$ will blow up as $t\to\infty$ and some will vanish since it is a sum of exponentials with positive coefficients. The only exception to this rule would be if there is a solution of the F-term equations where the components of $\phi$ that grow precisely vanish. This would imply that there are non-trivial subspaces $S_k\subset\C^{N_k}$ such that
\begin{equation}
 \Phi_j^{(k)}S_k\subset S_k,
 \qquad
 Z^{(k\ell)}S_\ell\subset S_k
 \quad\text{for all }j,k,\ell.
 \label{eq:preserved-subspaces}
\end{equation}
Define the dimensions $m_k=\dim S_k$, such that $m_1$, $m_2$ and $m_3$ are either $0$ or $1$ and $0\leq m_4\leq N$.
Similar to the approach used to prove $X_a^{(k)}=0$, restricting the F-equations associated with the $i$-$4$ pair
to these subspaces and taking traces gives
\begin{equation}
\Tr_{S_4}\!\big(Z^{(4i)}Z^{(i4)}\big)
 =-c^{(i4)}m_4,\qquad
\Tr_{S_i}\!\big(Z^{(i4)}Z^{(4i)}\big)
 =-Nc^{(i4)}m_i.
 \label{eq:restricted-F-traces}
\end{equation}
All commutators have zero trace.
The two traces on the LHS are equal. This is easy to see by writing the two traces explicitly. Therefore
$
 c^{(i4)}(m_4-Nm_i)=0.
$
The assumption $c^{(i4)}\neq0$ implies
\begin{equation}
 m_4=Nm_1=Nm_2=Nm_3.
\end{equation}
The only possibilities are that all subspaces vanish or that
all are the full spaces. Therefore an intermediate situation where $\phi$ is such that the growing solutions of $\norm{g\cdot \phi}^2$ are absent is not possible. This argument already hints at why cases with general $(N_1,N_2,N_3,N_4)$ can be more subtle depending on arithmetic properties.

\subsection*{Final form of the BPS equations}

As preparation for the explicit counting carried out in the rest of the paper, we will further simplify the F-term equations as much as possible in the following way. We first remove the six free complex coordinates associated with
the shift symmetries by introducing
\begin{equation}
 \begin{gathered}
 A_i=\Phi_i^{(4)}-\Phi_i^{(i)}I_N,\qquad i=1,2,3,\\
 \alpha_{12}=\Phi_3^{(1)}-\Phi_3^{(2)},\qquad
 \alpha_{23}=\Phi_1^{(2)}-\Phi_1^{(3)},\qquad
 \alpha_{31}=\Phi_2^{(3)}-\Phi_2^{(1)}.
 \end{gathered}
 \label{eq:invariant-adjoints}
\end{equation}
These three complex $N\times N$ matrices and three scalars
parametrize the adjoint fields modulo the shifts. This removes all shifts redundancies since we already determined the vector multiplets have no vev. For the fields representing open strings between distinct D-branes introduce the notation
\begin{equation}
 z_{ij}=Z^{(ij)}\quad(1\leq i\ne j\leq3),\qquad
 u_i=Z^{(4i)}\in\C^N,\qquad
 v_i=Z^{(i4)}\in(\C^N)^*.
 \label{eq:reduced-arrows}
\end{equation}
We assume that $c^{(12)},c^{(23)},c^{(31)}$ are all nonzero throughout. 
The $\alpha_{ij}$ appear only as Lagrange multipliers
\begin{equation}
 \frac{W}{\sqrt2}
 \supset
 \alpha_{12}\bigl(z_{12}z_{21}+c^{(12)}\bigr)
 +\alpha_{23}\bigl(z_{23}z_{32}+c^{(23)}\bigr)
 +\alpha_{31}\bigl(z_{31}z_{13}+c^{(31)}\bigr).
\end{equation}
They can be easily integrated out. In particular, they imply that all the open string fields between separate D2-branes are non-zero. The BPS vacua therefore consist of bound states of all three D2-branes while the connection to the D6-brane stack depends on the remaining details. Using that the $z_{ij}$ are non-zero we can use the complexified gauge transformations to fix
\begin{equation}
 \begin{gathered}
 z_{12}=z_{23}=1,\qquad
 z_{21}=-c^{(12)},\qquad z_{32}=-c^{(23)},\\
 z_{31}=t,\qquad z_{13}=-\frac{c^{(31)}}{t},
 \qquad t\neq0.
 \end{gathered}
 \label{eq:scalar-gauge-chart}
\end{equation}
This choice sets $g_1=g_2=g_3$, and the common D2 factor is removed by the quotient by $\C^\ast_{\rm diag}$. The remaining complex gauge group is therefore $\mathrm{GL}(N,\C)$,
acting by conjugation on the $A_i$ and in the fundamental
and dual representations on $u_i$ and $v_i$, while the product $z_{12}z_{23}z_{31}=t$ is gauge invariant.

Substituting these relations into the superpotential gives the gauge-fixed superpotential $W_{\rm gf}$,
\begin{equation}
 \begin{aligned}
 \frac{W_{\rm gf}}{\sqrt2}={}&
 t-\frac{c^{(12)}c^{(23)}c^{(31)}}{t}
 -\Tr A_1[A_2,A_3]
 -\sum_{i=1}^3v_iA_iu_i\\
 &-\sum_{i=1}^3c^{(i4)}\Tr A_i
 +v_2u_1-c^{(12)}v_1u_2\\
 &+\frac{c^{(31)}}{t}v_3u_1+tv_1u_3
 +v_3u_2-c^{(23)}v_2u_3.
 \end{aligned}
 \label{eq:reduced-W-unnormalized}
\end{equation}
This superpotential can be simplified further by rescaling all fields and sources. Introduce the combination
$
 \Lambda^2=c^{(12)}c^{(23)}c^{(31)},
$ and choose a square root branch. Make the invertible holomorphic change of variables
\begin{equation}
 \begin{gathered}
 t=\Lambda\widehat t,\qquad
 A_1=\frac{\Lambda}{c^{(23)}}\widehat A_1,\qquad
 A_2=\frac{\Lambda}{c^{(31)}}\widehat A_2,\qquad
 A_3=\frac{\Lambda}{c^{(12)}}\widehat A_3,\\
 u_1=\widehat u_1,\qquad
 u_2=\frac{\Lambda}{c^{(12)}c^{(23)}}\widehat u_2,\qquad
 u_3=\frac{1}{c^{(23)}}\widehat u_3,\\
 v_1=c^{(23)}\widehat v_1,\qquad
 v_2=\Lambda\widehat v_2,\qquad
 v_3=c^{(12)}c^{(23)}\widehat v_3.
 \end{gathered}
 \label{eq:normalizing-redefinitions}
\end{equation}
The remaining background parameters become
\begin{equation}
 \widehat c^{(14)}=\frac{c^{(14)}}{c^{(23)}},\qquad
 \widehat c^{(24)}=\frac{c^{(24)}}{c^{(31)}},\qquad
 \widehat c^{(34)}=\frac{c^{(34)}}{c^{(12)}}.
 \label{eq:normalized-sources}
\end{equation}
With these definitions, every term in
\eqref{eq:reduced-W-unnormalized} acquires the same overall
factor $\Lambda$. Thus
$W_{\rm gf}=\sqrt2\Lambda\,\widehat W$, where $\widehat W$
has the form obtained by setting the three D2-D2 coefficients
to one. Both the invertible field redefinition and removal of
the nonzero overall factor preserve the critical-point equations.
This rescaling applies only to the superpotential and is not necessarily preserved by the D-terms of kinetic gauge terms. Since we already took care of both this is irrelevant. From now on, we omit the hats and write $W$ for the normalized
superpotential,
\begin{equation}
 W={}t-\frac1t
 -\Tr A_1[A_2,A_3]
 -\sum_{i=1}^3v_iA_iu_i
 -\sum_{i=1}^3c^{(i4)}\Tr A_i+\sum_{i,j}v_iK_{ij}(t)u_j
 \label{eq:normalized-superpotential}
\end{equation}
where
\begin{equation}\label{eq:Kdefinition}
 K(t)=\begin{pmatrix}0&-1&t\\1&0&-1\\1/t&1&0\end{pmatrix}.
\end{equation}
This is the form of the potential that we will use in the next section. All subsequent fields and parameters are understood in these
normalized variables. The only explicit background parameters
remaining in the F-equations are the three normalized
$c^{(i4)}$. In the next section we will count the number of critical points of $W$ in the regime of small but non-zero $c^{(i4)}$.

\section{BPS configurations for the $(1,1,1,N)$ system}\label{sec:counting}

We first solve the BPS equations explicitly at $c^{(i4)}=0$ in section~\ref{sec:ZeroSourceSolutions} and determine the resulting moduli space. This problem becomes tractable once we restrict to configurations that can be extended to non-zero $c^{(i4)}$. In section~\ref{sec:NonzeroSource}, we turn on $c^{(i4)}$ perturbatively, show that the deformation lifts the continuous moduli, and count the resulting isolated vacua. For completeness, section~\ref{sec:numerical} compares the perturbative solutions we constructed and the exact solutions from numerical methods, and confirms their validity at finite $\delta$.

\subsection{Configurations at $c^{(i4)}=0$}
\label{sec:ZeroSourceSolutions}

The goal is to find the critical points of the superpotential \eqref{eq:normalized-superpotential} with the sources $c^{(i4)}$ set to zero. We begin by writing the F-term equations explicitly. Varying the three adjoint matrices gives
\begin{equation}
 \begin{aligned}
 [A_2,A_3]+u_1v_1&=0,\\
 [A_3,A_1]+u_2v_2&=0,\\
 [A_1,A_2]+u_3v_3&=0.
 \end{aligned}
 \label{eq:F-adjoints-c0}
\end{equation}
Varying the rows $v_i$ and columns $u_i$ gives, respectively,
\begin{equation}
 \begin{aligned}
 A_1u_1&=-u_2+t u_3,
 &\qquad v_1A_1&=v_2+\frac1t v_3,\\
 A_2u_2&=u_1-u_3,
 &v_2A_2&=-v_1+v_3,\\
 A_3u_3&=\frac1t u_1+u_2,
 &v_3A_3&=t v_1-v_2.
 \end{aligned}
 \label{eq:F-framing-c0}
\end{equation}
Finally, varying $t\neq0$ gives
\begin{equation}
 1+\frac1{t^2}-\frac1{t^2}v_3u_1+v_1u_3=0.
 \label{eq:F-t-c0}
\end{equation}
When we turn on $c^{(i4)}$, the only modification is to the equations \eqref{eq:F-adjoints-c0} that arise from varying the adjoint fields. We will see this small change can have crucial consequences for the counting.

We begin with the simplest case, $N=0$. With no D6-branes, only the D2-brane sector remains, and the superpotential reduces to $W=t-1/t$. Its two critical points, $t=\pm\i$, give $d_{N=0}=2$.

For $N\geq 1$, we must solve the matrix equations above. This problem is a special case of a more general problem that appears in the mathematical literature such as \cite{King94, CB01,Przezdziecki:2016}. Instead of relying on those references, here we present a direct argument tailored to our model.

We first characterize families of solutions at $c^{(i4)}\neq0$ that admit a finite limit as the sources vanish. By choosing appropriate gauge transformations in $G_\C$, we can arrange for the limiting configuration to decompose into an $R$-dimensional block, we call the ``core'', and a direct sum of scalar blocks, namely
\begin{equation}\label{eq:zero-source-blocks}
\begin{gathered}
 A_i^{(0)}=\begin{pmatrix}
 A_i^{\rm core}&0&\cdots&0\\
 0&a_{i,1}I_{n_1}&\cdots&0\\
 \vdots&\vdots&\ddots&\vdots\\
 0&0&\cdots&a_{i,S}I_{n_S}
 \end{pmatrix},\\
 u_i^{(0)}=\begin{pmatrix}u_i^{\rm core}\\0\\\vdots\\0\end{pmatrix},
 \qquad
 v_i^{(0)}=\begin{pmatrix}v_i^{\rm core}&0&\cdots&0\end{pmatrix},
 \qquad t^{(0)}=t_{\rm core},\\
 V=V_{\rm core}\oplus\bigoplus_{s=1}^{S}\C^{n_s},\qquad n=N-R=\sum_{s=1}^S n_s.
\end{gathered}
\end{equation}
The index $s=1,\ldots,S$ labels a distinct triple $(a_{1,s},a_{2,s},a_{3,s})\in\C^3$, with multiplicity $n_s$. We will show below that the core solutions occur in complex-conjugate pairs, each pair uniquely specified by a nonnegative integer $r$, with core dimension $R=r(r+1)$. For a fixed core, the remaining moduli are the unordered scalar positions, such that
\begin{equation}\label{eq:zero-source-moduli}
 \operatorname{Sym}^{n}(\C^3)=(\C^3)^n/S_n,\qquad n=N-R.
\end{equation}
For $n>0$, the vacua form a continuous family and cannot be counted as isolated points at $c^{(i4)}=0$. In the next section, we turn on $c^{(i4)}$ perturbatively, show that this deformation lifts the continuous moduli, and count the resulting isolated vacua.

\smallskip

Let us first argue why any relevant solution can be put in this form. Section~2 shows that, when the sources are nonzero, there are no proper nonzero subspaces preserved by the $A_i$ that either contain all the $u_i$'s or are annihilated by all the $v_i$'s.
The argument fails when $c^{(i4)}=0$ so such subspaces can appear in a finite zero-source limit.
Choose bases adapted to the invariant subspaces, putting the $A_i$ in block upper-triangular form with corresponding blocks for the $u_i$'s
and $v_i$'s, and repeat until no diagonal block admits a further invariant subspace of this kind. Since the $z_{ij}$ are nonzero, the three one-dimensional D2 spaces must remain in the same diagonal block, which we called the core above. Every other diagonal block has vanishing $u_i$'s and $v_i$'s, so its three $A_i$ commute according to the F-equations. They therefore have a common eigenvector and can be simultaneously diagonalized as in \eqref{eq:zero-source-blocks}. Finally, complex gauge rescalings of the coordinates can be chosen in a way that the blocks above the diagonal mixing the core and the scalars vanish in the $c^{(i4)} \to 0$ limit\footnote{For example, consider a matrix $ A(\delta)=\Big(\,\begin{matrix}a&b\\ \delta^2c&d\end{matrix}\,\Big),
$ where $a$ would play the role of the core and $d$ the role of the scalar block. If we define $g(\delta) = \text{diag}(1, \delta^{-1})$ a simple calculation shows that $g(\delta)A(\delta)g(\delta)^{-1} \to \text{diag}(a,d)$ as $\delta\to 0$. The transformation is invertible for every nonzero $\delta$, and the same argument applies to larger blocks.}. This decomposition applies to finite limits of solutions at nonzero $c^{(i4)}$, while arbitrary solutions at $c^{(i4)}=0$ need not have this form. In the next section, we will show that extension to generic nonzero sources further restricts each scalar triple $(a_{1,s},a_{2,s},a_{3,s})$ to one of six possible values. For now, we leave these triplets arbitrary.

\smallskip

We now focus on the core block, which independently satisfies the same F-term equations with $N$ replaced by $R$. Restricting to this block and suppressing the core superscript, we introduce the matrix resolvents
\begin{equation}
 R_i(z)=(zI-A_i)^{-1},
 \qquad i=1,2,3,
 \label{eq:core-resolvents}
\end{equation}
and the following six rational functions:
\begin{equation}
 \begin{aligned}
 p_1(z)&=z-t^{-1}-z\,v_2R_1(z)u_3,
 &q_1(z)&=z-t-z\,v_3R_1(z)u_2,\\
 p_2(z)&=z+t+tz\,v_1R_2(z)u_3,
 &q_2(z)&=z+t^{-1}-\frac{z}{t}\,v_3R_2(z)u_1,\\
 p_3(z)&=z-t^{-1}-z\,v_1R_3(z)u_2,
 &q_3(z)&=z-t-z\,v_2R_3(z)u_1.
 \end{aligned}
 \label{eq:core-six-functions}
\end{equation}
After some tedious algebra, using the F-equations to evaluate products and derivatives of
these contractions gives
\begin{equation}
 p_iq_i=z^2+1,\qquad
 \frac{p_i'}{p_i}=\frac{z+\kappa}{z^2+1},
 \qquad
 \frac{q_i'}{q_i}=\frac{z-\kappa}{z^2+1},
 \qquad i=1,2,3,
 \label{eq:core-pq-differential}
\end{equation}
where the same constant
\begin{equation}
\kappa=v_1u_2+t^{-1}=v_2u_3+t^{-1}
 \label{eq:core-kappa}
\end{equation}
appears for all three adjoints.
The fact that the RHS in these equations are the same for all $i$ immediately implies the ratios $p_i/p_j$ and $q_i/q_j$ are constant. In fact in \eqref{eq:core-six-functions} the extra powers of $t$ were chosen so that this property holds.
Moreover, at large $z$ all functions go as $p_i(z) \sim z$ and $q_i (z) \sim z$ which fixes the constant ratios to one. Therefore
\begin{equation}
 p_1=p_2=p_3\equiv p,\qquad
 q_1=q_2=q_3\equiv q.
 \label{eq:core-pq-equality}
\end{equation}
Now we can go back to the differential equations for $p(z)$ and $q(z)$ and solve them
\beq
p(z) = (z-\i)^{\frac{1}{2}+\frac{\kappa}{2\i}}(z+\i)^{\frac{1}{2}-\frac{\kappa}{2\i}},\qquad q(z)=(z-\i)^{\frac{1}{2}-\frac{\kappa}{2\i}}(z+\i)^{\frac{1}{2}+\frac{\kappa}{2\i}},
\eeq
the integration constant was fixed by the requirement that $p,q \sim z$ at large $z$. Going back to the definition in \eqref{eq:core-six-functions}, it is clear that these functions are rational. Therefore the order of a zero or a pole has to be an integer. This quantizes the value of $\kappa$ in the solution such that 
\beq
\kappa = \i (2r+1),\qquad r\in \mathbb{Z}. 
\eeq
This can be used to solve once and for all for $t$. By definition, using $p_2(z)$ form, we have the relationship $p(0)=t$ and using the solution above we can conclude
\beq
t = \i (-1)^{r+1}.
\eeq
From now on we will restrict to $r\geq0$. The normalized F-equations
are invariant under complex conjugation, which sends
$r\to-r-1$ and $t\to\bar t=-t$, and thus generates
the remaining branches with $r\leq-1$. This will make all solutions we find below doubly degenerate if parametrized in terms of $r\geq 0$. 

\smallskip

We can now argue that the core solution is unique up to conjugation. We can use the Taylor expansion of $p_i(z)$ and $q_i(z)$ to compute any contraction such as 
\beq
v_i A_k^n u_j,\qquad \text{for $i$,$j$,$k$ distinct}.
\eeq
With this information, together with the F-equations, we can compute any contraction
\beq
\label{eq:contraction}
v_i P(A_1,A_2,A_3) u_j,
\eeq
where $P(A_1,A_2,A_3)$ is an arbitrary polynomial in the matrices. Start with a contraction such that $i \neq j$. In that case, we want to remove any appearance of $A_k$ where either $k=i$ or $k=j$. If $k=i$ we can use the commutator F-equations \eqref{eq:F-adjoints-c0} to move $A_k$ to the right and act on $u_i$. The leftover term $u_a v_b$ can be used to split the contraction into shorter ones.  At the end of the process, we can use \eqref{eq:F-framing-c0} to exchange $A_i u_i$ by a linear combination of $u_j$. The only case left to consider is when we encounter $v_i P(A_1,A_2,A_3) u_i$. Take $i=1$ for concreteness. In this case we can again eliminate $A_1$ by making it act on $u_1$ or $v_1$ and reduce the contraction to $v_1 \tilde{P}(A_2,A_3) u_1$. It is not too hard to show that any such contraction vanishes by taking traces of the F-equations and using induction\footnote{For a fixed $i,j,k$ write $X=A_j$, $Y=A_k$, $u=u_i$, and $v=v_i$, with $(i,j,k)$ cyclic, so the F-equations give $[X,Y]=-uv$.
Taking the trace gives $vu=0$. We can prove this by induction on the length of a word in $X,Y$. Exchanging adjacent $X,Y$ inside $vP(X,Y)u$ produces a product of shorter contractions, which vanish by induction. Thus the ordering does not matter, and $0=\Tr[X,X^aY^{b+1}]=-(b+1)vX^aY^bu$. This proves that every such contraction vanishes, and hence $v_iP(A_j,A_k)u_i=0$ for any polynomial $P$.}. The last step is to construct a basis of vectors on $V_{\text{core}}$ out of polynomials of $A_1$, $A_2$ and $A_3$ acting on the $u_i$'s and similarly for the $v_i$'s. This is possible given the way the core was constructed. If that is the case we can apply this procedure to compute the matrix elements of the adjoint fields as well as the vectors $u_i$ and $v_j$.  This way we construct explicitly a unique solution up to conjugation.
 
We have shown that the relevant limiting configurations take the form \eqref{eq:zero-source-blocks}, with cores occurring in complex-conjugate pairs labeled by a nonnegative integer $r$ and found $t$. It remains to establish that their dimension is $R=r(r+1)$. This can be extracted from \cite{Przezdziecki:2016} after comparing our equations with their notation. Instead of doing that, we give an intuitive understand of where it comes from. 

Define $J_\ell$ such that $(J_\ell)_{a,a+1}=1$ and all other entries vanish. A Jordan block $\lambda I_\ell+J_\ell$ in $A_i$, with
$\lambda=\pm\i$, produces a pole of order $\ell$ in the
corresponding function $p_i$ or $q_i$, assuming that the highest-order pole of
the resolvent survives the contraction. Under this assumption, we can construct a solution at level $r+1$ from a solution at level $r$ by appending two Jordan blocks $(\i I_{r+1} + J_{r+1} ) \oplus (-\i I_{r+1} + J_{r+1} )$. This raises the dimension by $2(r+1)$ and therefore 
\beq
R(r+1) = R(r) + 2(r+1) ,\qquad R=r(r+1).\label{eq:core-dimension}
\eeq
The initial condition for the recursion can be obtained by the explicit $r=1$ solution 
\beq
t=\i,\qquad
A_1=\begin{pmatrix}\i&0\\0&-\i\end{pmatrix},\quad
A_2=\begin{pmatrix}\i&0\\-2&-\i\end{pmatrix},\quad
A_3=\begin{pmatrix}-\i&2\\0&\i\end{pmatrix}.
\eeq
The columns are 
\beq
u_1=\begin{pmatrix}\i\\-1\end{pmatrix},\qquad
u_2=\begin{pmatrix}1\\0\end{pmatrix},\qquad
u_3=\begin{pmatrix}0\\1\end{pmatrix},
\eeq
and the rows are 
\beq
v_1=(4\i,-4),\qquad
v_2=(0,4\i),\qquad
v_3=(-4\i,0).
\eeq
The second solution is obtained by complex conjugation. It is a simple exercise to verify this solves the equations.

This construction motivates the following property for the adjoint matrices 
\begin{equation}
 A_i\sim
 \bigoplus_{\ell=1}^{r}(\i I_\ell+J_\ell)
 \oplus
 \bigoplus_{\ell=1}^{r}(-\i I_\ell+J_\ell),
 \label{eq:core-jordan-form-derived}
\end{equation}
A more formal proof of this fact can be obtained as a special case of the construction in \cite{Przezdziecki:2016}. The Jordan decomposition also provides a starting point for constructing the core explicitly. Choose a basis in which $A_1$ has this form, with blocks ordered as $V_{+r},\ldots,V_{+1},V_{-r},\ldots,V_{-1}$, and set $(u_2)_a=\delta_{ar}$ and $(u_3)_a=\delta_{a,R/2+r}$. The remaining fields can then be determined from the F-term equations.

\subsection{Configurations at $c^{(i4)}\neq0$}
\label{sec:NonzeroSource}

We now turn on the sources $c^{(i4)}$. As noted in Section~\ref{sec:ZeroSourceSolutions}, they enter only the adjoint F-equations \eqref{eq:F-adjoints-c0} through a constant term $c^{(i4)}I_N$. We parametrize these sources as
\begin{equation}\label{eq:relative-source}
 c^{(i4)}=c_i\delta^2,\qquad i=1,2,3,
\end{equation}
with fixed generic $c_i$ and a parameter $\delta$.
In Section~\ref{sec:exact-expansion}, we rewrite the superpotential in terms of the fluctuations around the core plus scalars configuration \eqref{eq:zero-source-blocks}. After using gauge symmetry to make $A_1$ block diagonal, we solve the linear equations for the off-diagonal blocks of $A_2$ and $A_3$. Substituting the solutions back produces a reduced superpotential involving the core, the rows and columns, and the diagonal blocks of $A_i$ only. For every scalar block, one of the equations implies an exact constraint of the Calogero-Moser type. The remaining equations are coupled and nonlinear, and for general $N$ we do not have a closed-form solution.

We therefore proceed with small $\delta$ and solve the F-term equations order by order in Section~\ref{sec:small-delta}. At order $\delta$, we find a cubic surface for the scalar values in \eqref{eq:zero-source-moduli}, on which the D2-D6 strings on each scalar block develop a massless mode. At order $\delta^2$, the source term induces a linear potential on this surface, its critical points select six distinct scalar values and therefore lift the moduli \eqref{eq:zero-source-moduli}. Some fields remain light and form a Calogero-Moser space for each scalar block. At order $\delta^3$, we identify a quadratic potential on each Calogero-Moser space, whose critical points are isolated, nondegenerate, and labeled by partitions of $n_s$.

\subsubsection{Exact superpotential and its reduction}
\label{sec:exact-expansion}

Allowing every block of the zero-source background \eqref{eq:zero-source-blocks} to fluctuate,
\begin{equation}\label{eq:FluctuationAnsatz}
\begin{gathered}
A_i=
\begin{pmatrix}
A_i^{\rm core}+\Delta A_i^{\mathrm{cc}}
 &\Delta A_i^{\mathrm{c}1}&\cdots&\Delta A_i^{\mathrm{c}S}\\
\Delta A_i^{1\mathrm c}
 &a_{i,1}I_{n_1}+\Delta A_i^{11}&\cdots&\Delta A_i^{1S}\\
\vdots&\vdots&\ddots&\vdots\\
\Delta A_i^{S\mathrm c}
 &\Delta A_i^{S1}&\cdots&a_{i,S}I_{n_S}+\Delta A_i^{SS}
\end{pmatrix},\\
u_i=\begin{pmatrix}u_i^{\rm core}+\Delta u_i^{\mathrm c}\\
\Delta u_i^1\\ \vdots\\ \Delta u_i^S\end{pmatrix},
\qquad
v_i=\begin{pmatrix}v_i^{\rm core}+\Delta v_i^{\mathrm c}&
\Delta v_i^1&\cdots&\Delta v_i^S\end{pmatrix},
\qquad t=t_{\rm core}+\Delta t .
\end{gathered}
\end{equation}
The label $\mathrm c$ denotes the core of dimension $R$, and $s=1,\ldots,S$ labels the scalar blocks of dimensions $n_s$. We call $\Delta A_i^{\rm cc}$, $\Delta u_i^{\rm c}$, $\Delta v_i^{\rm c}$ and $\Delta t$ the core fluctuations, and $\Delta A_i^{ss}$ the scalar block adjoints. The off-diagonal adjoints are of two kinds. The core-block adjoints $\Delta A_i^{{\rm c}s}$ and $\Delta A_i^{s{\rm c}}$ have sizes $R\times n_s$ and $n_s\times R$. The block-block adjoints $\Delta A_i^{\ell m}$, with $\ell\neq m$, have size $n_\ell\times n_m$. Finally, the columns $\Delta u_i^s$ and rows $\Delta v_i^s$ are the D2${}_i$-D6 strings ending on block $s$. These fluctuations are not assumed to be small, we expand them in $\delta$ only in Section~\ref{sec:small-delta}.

Substituting \eqref{eq:FluctuationAnsatz} into \eqref{eq:normalized-superpotential} gives the exact expansion
\begin{equation}\label{app:zero:expansion}
\begin{aligned}
 W(\delta)={}&W_0+\mathcal Q(\Delta\phi,\Delta\phi)+\mathcal R(\Delta\phi)\\
 &-\delta^2\sum_i c_i\left[
 \Tr_R(A_i^{\rm core}+\Delta A_i^{\mathrm{cc}})
 +\sum_s\bigl(n_sa_{i,s}+\Tr_{n_s}\Delta A_i^{ss}\bigr)\right],\\
 \mathcal Q(\Delta\phi,\Delta\phi)={}&\mathcal Q_{\rm core}
       +\sum_s \mathcal Q_{\mathrm cs}+\sum_{\ell<m}\mathcal Q_{\ell m}.
\end{aligned}
\end{equation}
It consists of the background value $W_0$, a quadratic form $\mathcal Q$, a remainder $\mathcal R$ that collects all cubic and higher terms, and the source term which is linear in the fields and proportional to $\delta^2$. The quadratic form splits into three parts. $\mathcal Q_{\rm core}$ contains the core fluctuations. For each scalar block, $\mathcal Q_{{\rm c}s}$ contains its columns and rows and its core-block adjoints. For each pair of scalar blocks, $\mathcal Q_{\ell m}$ contains the block-block adjoints. There is no quadratic term in diagonal blocks $\Delta A_i^{ss}$ since the background in a scalar block is proportional to the identity. All these pieces are given explicitly in \eqref{app:zero:core-quadratic}-\eqref{app:zero:remainder}. The F-term equations in terms of fluctuations are collected in \eqref{eq:exact-fluctuation-adj}-\eqref{eq:exact-fluctuation-t}.

\subsubsection*{Block gauge and eliminating the off-diagonal adjoints}
\label{sec:exact-block-equations}

BPS configurations are counted modulo the complexified gauge group $G_\C$ as in \eqref{eq:complex-gauge-quotient}. After carrying out the simplifications and choices that led to \eqref{eq:normalized-superpotential}, the residual group is $\GL(N,\C)$, acting as
$$
 A_i\longmapsto gA_ig^{-1},\qquad
 u_i\longmapsto gu_i,\qquad
 v_i\longmapsto v_ig^{-1},\qquad
 t\longmapsto t,\qquad g\in\GL(N,\C).
$$
We can decompose $g$ into blocks in the same way as the adjoints $A_i$ in \eqref{eq:FluctuationAnsatz}, then use its off-diagonal blocks to set the off-diagonal blocks of one of the three adjoints to zero. We choose $A_1$, which gives the block gauge
\begin{equation}\label{eq:exact-block-gauge}
\begin{gathered}
 \Delta A_1^{\mathrm cs}=\Delta A_1^{s\mathrm c}=0,\qquad
 \Delta A_1^{\ell m}=0\quad(\ell\ne m),\\
 A_1=\operatorname{diag}\!\left(
 A_1^{\rm core}+\Delta A_1^{\mathrm{cc}},\,
 a_{1,1}I_{n_1}+\Delta A_1^{11},\,
 a_{1,2}I_{n_2}+\Delta A_1^{22},\,\ldots\right).
\end{gathered}
\end{equation}
The blocks are organized according to scalar values, for example $a_{1,\ell} \neq a_{1,m} $ for $\ell \neq m$. This gauge condition is valid as long as all these values together with core block eigenvalues are distinct from each other, which we confirm below. The remaining gauge symmetries are block-diagonal
$$
 g=\operatorname{diag}(g_{\mathrm c},g_1,\ldots,g_S),\qquad
 g_{\mathrm c}\in\GL(R,\C),\quad g_s\in\GL(n_s,\C).
$$
The core factor $\GL(R,\C)$ acts on the core fluctuations, and the factor $\GL(n_s,\C)$ acts on the Calogero-Moser data of block $s$, see \eqref{eq:reduced-exact-CM-gauge}.

This choice of gauge makes the equations for the remaining off-diagonal adjoints linear. Write $A_i^{\rm cc}=A_i^{\rm core}+\Delta A_i^{\rm cc}$ and $A_i^{ss}=a_{i,s}I_{n_s}+\Delta A_i^{ss}$ for the full diagonal block matrices. The ${\rm c}s$ and $s{\rm c}$ blocks of the second and third equations in \eqref{eq:exact-fluctuation-adj} are given by
\begin{equation}\label{eq:exact-core-sylvester}
\begin{aligned}
 A_1^{\rm cc}\Delta A_2^{\mathrm cs}-\Delta A_2^{\mathrm cs}A_1^{ss}
   &=-(u_3^{\rm core}+\Delta u_3^{\mathrm c})\Delta v_3^s,\\
 A_1^{\rm cc}\Delta A_3^{\mathrm cs}-\Delta A_3^{\mathrm cs}A_1^{ss}
   &=(u_2^{\rm core}+\Delta u_2^{\mathrm c})\Delta v_2^s,\\
 \Delta A_2^{s\mathrm c}A_1^{\rm cc}-A_1^{ss}\Delta A_2^{s\mathrm c}
   &=\Delta u_3^s(v_3^{\rm core}+\Delta v_3^{\mathrm c}),\\
 \Delta A_3^{s\mathrm c}A_1^{\rm cc}-A_1^{ss}\Delta A_3^{s\mathrm c}
   &=-\Delta u_2^s(v_2^{\rm core}+\Delta v_2^{\mathrm c}).
\end{aligned}
\end{equation}
When the spectra of $A_1^{\rm cc}$ and $A_1^{ss}$ are disjoint, each has a unique solution. Expanding $A_1^{ss}$ about $a_{1,s}I_{n_s}$ gives the solution as the series
\begin{equation}\label{eq:exact-core-connecting-series}
\begin{aligned}
 \Delta A_2^{\mathrm cs}
   &=-\sum_{m\ge0}(A_1^{\rm cc}-a_{1,s}I_R)^{-m-1}
      (u_3^{\rm core}+\Delta u_3^{\mathrm c})\Delta v_3^s(\Delta A_1^{ss})^m,\\
 \Delta A_3^{\mathrm cs}
   &=\sum_{m\ge0}(A_1^{\rm cc}-a_{1,s}I_R)^{-m-1}
      (u_2^{\rm core}+\Delta u_2^{\mathrm c})\Delta v_2^s(\Delta A_1^{ss})^m,\\
 \Delta A_2^{s\mathrm c}
   &=\sum_{m\ge0}(\Delta A_1^{ss})^m\Delta u_3^s
      (v_3^{\rm core}+\Delta v_3^{\mathrm c})(A_1^{\rm cc}-a_{1,s}I_R)^{-m-1},\\
 \Delta A_3^{s\mathrm c}
   &=-\sum_{m\ge0}(\Delta A_1^{ss})^m\Delta u_2^s
      (v_2^{\rm core}+\Delta v_2^{\mathrm c})(A_1^{\rm cc}-a_{1,s}I_R)^{-m-1},
\end{aligned}
\end{equation}
which converges near the background when $a_{1,s}$ are distinct from the eigenvalues of $A_1^{\rm cc}$. Similarly, for $\ell\neq m$, the block-block adjoint equations are
\begin{equation}\label{eq:exact-scalar-sylvester}
 A_1^{\ell\ell}\Delta A_2^{\ell m}-\Delta A_2^{\ell m}A_1^{mm}=-\Delta u_3^\ell\Delta v_3^m,\qquad
 A_1^{\ell\ell}\Delta A_3^{\ell m}-\Delta A_3^{\ell m}A_1^{mm}=\Delta u_2^\ell\Delta v_2^m,
\end{equation}
whose solutions for $a_{1,\ell}\neq a_{1,m}$ are given by
\begin{equation}
\begin{aligned}
 \Delta A_2^{\ell m}
   &=-\sum_{p,q\ge0}\binom{p+q}{p}(a_{1,\ell}-a_{1,m})^{-p-q-1}
      (-\Delta A_1^{\ell\ell})^p\,\Delta u_3^\ell\Delta v_3^m\,(\Delta A_1^{mm})^q,\\
 \Delta A_3^{\ell m}
   &=\sum_{p,q\ge0}\binom{p+q}{p}(a_{1,\ell}-a_{1,m})^{-p-q-1}
      (-\Delta A_1^{\ell\ell})^p\,\Delta u_2^\ell\Delta v_2^m\,(\Delta A_1^{mm})^q.
\end{aligned}
\label{eq:DeltaA2DeltaA3}
\end{equation}

All off-diagonal adjoints are therefore determined by the diagonal blocks of $A_1$ and by the columns and rows. The remaining off-diagonal equations in \eqref{eq:exact-fluctuation-adj} are the equations of the gauge-fixed blocks of $A_1$. We show they are satisfied in Appendix~\ref{app:exact-W}.

\subsubsection*{Reduced superpotential}
\label{sec:reduced-superpotential}

In the block gauge, integrating out these off-diagonal adjoints by substituting the solutions back into $W$, we arrive at a reduced superpotential
\begin{equation}\label{eq:exact-reduced-superpotential}
\begin{aligned}
 W_{\rm red}={}&W_{\rm core}
 -\sum_s\Tr_{n_s}\Delta A_1^{ss}[\Delta A_2^{ss},\Delta A_3^{ss}]
 -\delta^2\sum_{i,s}c_i\Tr_{n_s}\left(a_{i,s}I_{n_s}+\Delta A_i^{ss}\right)\\
 &+\sum_s\Delta\mathbf v_sM_s(A^{ss})\Delta\mathbf u_s
 -\sum_{\ell\ne m}\Delta v_2^\ell\,\Delta A_2^{\ell m}\,\Delta u_2^m,
\end{aligned}
\end{equation}
whose critical points are in one-to-one correspondence with those of $W$. The derivation of \eqref{eq:exact-reduced-superpotential} is given in Appendix~\ref{app:exact-W}.
$W_{\rm core}$ is the superpotential \eqref{eq:normalized-superpotential} restricted to the core block. It depends on the core fluctuations $\Delta A_i^{\rm cc}$, $\Delta u_i^{\rm c}$, $\Delta v_i^{\rm c}$ and $\Delta t$, and includes $t-1/t$ and the core source term $-\delta^2\sum_ic_i\Tr_RA_i^{\rm cc}$. Its quadratic part is $\mathcal Q_{\rm core}$. Appendix~\ref{app:core-mass} shows that all core fluctuations are massive once the $\GL(R,\C)$ gauge directions are removed. The core couples to the scalar blocks only through the functions $G$ and $H$ in the matrix $M_s(A^{ss})$ defined below.

The second and third terms in the first line are the cubic interaction and the source term of the adjoints in each scalar block. For the first term in the second line, we collect the columns and rows of block $s$ into $\Delta\mathbf u_s=(\Delta u_1^s,\Delta u_2^s,\Delta u_3^s)^{\mathsf T}$ and $\Delta\mathbf v_s=(\Delta v_1^s,\Delta v_2^s,\Delta v_3^s)$. They couple through the $3n_s\times3n_s$ matrix
\begin{equation}\label{eq:exact-scalar-matrix}
 M_s(A^{ss})=
 \begin{pmatrix}
 -A_1^{ss}&-I_{n_s}&tI_{n_s}\\
 I_{n_s}&-A_2^{ss}&-I_{n_s}-G(A_1^{ss})\\
 t^{-1}I_{n_s}&I_{n_s}+H(A_1^{ss})&-A_3^{ss}
 \end{pmatrix},
\end{equation}
where the core enters through the resolvent contractions
\begin{equation}\label{eq:exact-core-resolvent-functions}
\begin{aligned}
 G(z)&=(v_3^{\rm core}+\Delta v_3^{\mathrm c})
 (A_1^{\rm cc}-zI_R)^{-1}
 (u_2^{\rm core}+\Delta u_2^{\mathrm c}),\\
 H(z)&=(v_2^{\rm core}+\Delta v_2^{\mathrm c})
 (A_1^{\rm cc}-zI_R)^{-1}
 (u_3^{\rm core}+\Delta u_3^{\mathrm c}).
\end{aligned}
\end{equation}
A function of the matrix $A_1^{ss}$ is defined by its Taylor series about $a_{1,s}$, for example,
\begin{equation}\label{eq:exact-resolvent-matrix-evaluation}
\begin{aligned}
 H(A_1^{ss})={}&\sum_{m\ge0}
 \Bigl[(v_2^{\rm core}+\Delta v_2^{\mathrm c})
 (A_1^{\rm cc}-a_{1,s}I_R)^{-m-1}
 (u_3^{\rm core}+\Delta u_3^{\mathrm c})\Bigr]
 (\Delta A_1^{ss})^m.
\end{aligned}
\end{equation}
The background value of $M_s(A^{ss})$ is $M_s(a_{i,s})$, which we write explicitly later in \eqref{eq:first-uv-quadratic}.

The matrix $M_s(A^{ss})$ is the mass matrix of the D2-D6 strings ending on block $s$. Its diagonal entries $-A_i^{ss}$ are the Wilson lines of D6-brane on block $s$ relative to D2-brane wrapping $T_i^2$, inherited from \eqref{eq:W-PhiZZ}. The off-diagonal entries $\pm1$ and $t^{\pm1}$ come from the cubic couplings \eqref{eq:W-triangles}. With the D2-D2 strings fixed by \eqref{eq:scalar-gauge-chart}, each triangle containing a D2-D2 string becomes a bilinear coupling between D2-D6 strings. The core corrects the $(2,3)$ and $(3,2)$ entries through $G$ and $H$, which are generated by integrating out the core-block adjoints.

We denote the last term of \eqref{eq:exact-reduced-superpotential} by $W_{\rm links}= -\sum_{\ell\ne m}\Delta v_2^\ell\,\Delta A_2^{\ell m}\,\Delta u_2^m$. It couples distinct scalar blocks through the block-block adjoints \eqref{eq:DeltaA2DeltaA3}. Taking in that expansion, it starts with a quartic interaction between D2-D6 strings
$$
 W_{\rm links}=\sum_{\ell\neq m}\frac{(\Delta v_2^\ell\Delta u_3^\ell)(\Delta v_3^m\Delta u_2^m)}{a_{1,\ell}-a_{1,m}}+\ldots,
$$
where the dots represent terms at higher degrees in the fluctuation fields.

\subsubsection*{Calogero-Moser constraint}
The reduced superpotential \eqref{eq:exact-reduced-superpotential} is linear in $A_2^{ss}$ and $A_3^{ss}$, so they act as Lagrange multipliers. Varying them gives the constraints
\begin{equation}\label{eq:reduced-exact-CM}
  \begin{aligned}
     [A_3^{ss},A_1^{ss}]
 +\Delta u_2^s\Delta v_2^s+\delta^2c_2I_{n_s}&=0,\\
[A_1^{ss},A_2^{ss}]
 +\Delta u_3^s\Delta v_3^s+\delta^2c_3I_{n_s}&=0.
  \end{aligned}
\end{equation}
Taking the traces
\begin{equation}\label{eq:reduced-exact-CM-trace}
 \Delta v_2^s\Delta u_2^s=-n_sc_2\delta^2,\qquad \Delta v_3^s\Delta u_3^s=-n_sc_3\delta^2,
\end{equation}
show that once $\delta\neq0$, the D2-D6 strings on every scalar block must condense, $\Delta v_i^s\Delta u_i^s\neq0$ for $i=2,3$, to balance the source induced by $c^{(i4)}$. In particular, the configuration $\Delta u_i^s=\Delta v_i^s=0$ which solves the zero-source equations is no longer allowed. 

Both equations in \eqref{eq:reduced-exact-CM} have the form of the Calogero-Moser constraint \cite{Wilson}. They are not independent since $A_1^{ss}$ appears in both. Modulo redisual $\GL(n_s,\C)$ gauge transformations,
\begin{equation}\label{eq:reduced-exact-CM-gauge}
 A_{1,2}^{ss}\longmapsto g_sA_{1,2}^{ss}g_s^{-1},\quad \Delta u_3^s\longmapsto g_s\Delta u_3^s,\quad
 \Delta v_3^s\longmapsto\Delta v_3^sg_s^{-1},
 \qquad g_s\in\GL(n_s,\C).
\end{equation}
the second constraint defines a Calogero-Moser space~\cite{Wilson}. It has complex dimension $2n_s$, since the constraint and the free $\GL(n_s,\C)$ action each remove $n_s^2$ out of $2n_s^2+2n_s$ coordinates in the two matrices, the column, and the row.
The first constraint, together with the remaining F-equations collected in Appendix~\ref{app:exact-W}, further determines which points on the Calogero–Moser space extend to BPS vacua.
These equations are coupled and highly, so we do not have a closed-form solution for general $N$. We therefore solve them perturbatively in $\delta$. In Section~\ref{sec:small-delta} we will see that the two constraints become identical and form a single Calogero-Moser space at order $\delta^2$.

\subsubsection{Small $\delta$ expansion}
\label{sec:small-delta}
Here we look for solutions that approach the zero-source backgrounds of Section~\ref{sec:ZeroSourceSolutions} as $\delta\to0$. The expansion determines which points of \eqref{eq:zero-source-moduli} extend to $\delta\neq0$ and how their degeneracy is lifted. The trace conditions \eqref{eq:reduced-exact-CM-trace} set the natural scale. In a gauge where columns and rows are balanced, both $\Delta u_i^s$ and $\Delta v_i^s$ are of order $\delta$. We therefore expand all fluctuations in integer powers of $\delta$ starting at this order,
\begin{equation}\label{eq:delta-field-expansion}
\begin{aligned}
 (\Delta A_i^{\mathrm{cc}},\Delta A_i^{ss})
 &=\sum_{m\ge1}\delta^m
   (A_i^{\mathrm{cc},(m)},A_i^{ss,(m)}),\\
 (\Delta u_i^{\mathrm c},\Delta u_i^s,
   \Delta v_i^{\mathrm c},\Delta v_i^s)
 &=\sum_{m\ge1}\delta^m
   (u_i^{\mathrm c,(m)},u_i^{s,(m)},
    v_i^{\mathrm c,(m)},v_i^{s,(m)}),\\
 \Delta t&=\sum_{m\ge1}\delta^m t^{(m)},
\end{aligned}
\end{equation}
where the superscript $(m)$ denotes the coefficient of $\delta^m$. The expansion series of off-diagonal adjoints $\Delta A_i^{\mathrm cs},\Delta A_i^{s\mathrm c}, \Delta A_i^{lm}$ follow from \eqref{eq:exact-core-connecting-series} and \eqref{eq:DeltaA2DeltaA3}. Inserting the expansion into \eqref{eq:exact-reduced-superpotential} produces an expansion of the reduced superpotential
\begin{equation}\label{eq:W-delta-expansion}
 W(\delta)=W_0+\sum_{m\ge 2} \delta^m\mathcal W^{(m)}.
\end{equation}
$\mathcal W^{(m)}$ depends on the background moduli $a_{i,s}$ and on the field coefficients of order up to $m-1$.
We can similarly expand the F-equation of a field $\phi$ evaluated on the expansion as a series in $\delta$,
\begin{equation}\label{eq:W-F-order-counting}
 \mathcal E_\phi \equiv \frac{\partial W}{\partial\phi}
   =\sum_{m\ge1}\delta^m\mathcal E_\phi^{(m)},\qquad
 \frac{\partial\mathcal W^{(m)}}{\partial\phi^{(m')}}
   =\mathcal E_\phi^{(m-m')},\qquad m\ge m'\ge0.
\end{equation}
Here $\phi^{(0)}$ refers to the background moduli $a_{i,s}$, which are the only zeroth-order data not fixed at $\delta=0$. The background satisfies $\mathcal E_\phi^{(0)}=0$. The order $\delta^m$ equations can be obtained by varying $\mathcal W^{(m+1)}$ with respect to the leading coefficients $\phi^{(1)}$. Derivatives with the same value of $m-m'$ reproduce the same equation.

\subsubsection*{Order $\delta$ equations}
The order $\delta$ equations follow from $\mathcal W^{(2)}$. At this order there is no coupling between the core and the scalar blocks, since every such coupling in \eqref{eq:exact-reduced-superpotential} is at least cubic in the fluctuations. For the core, they are the linear equations \eqref{eq:first-core-system}. Appendix~\ref{app:core-mass} proves that every solution of these equations is related to an infinitesimal $\GL(R,\C)$ transformation, with $t^{(1)}=0$. Fixing the residual gauge symmetry in core block, the leading perturbation of the core vanishes,
\begin{equation}\label{eq:core-leading-order}
 A_i^{\mathrm{cc},(1)}=0,\qquad
 u_i^{\mathrm c,(1)}=v_i^{\mathrm c,(1)}=0,\qquad t^{(1)}=0.
\end{equation}
In other words, all physical fluctuations in the core block are massive and therefore zero at linear order. It first responds at order $\delta^2$, where it is driven by $c_iI_R$ and through the core-block adjoints by the strings on the scalar blocks, see \eqref{eq:second-core-sources}. The same conclusion holds for the empty core $R=0$ where the only core fluctuation is $\Delta t$. $\mathcal Q_{\rm core}=-(\Delta t)^2/t_{\rm core}^3$ with $t_{\rm core}=\pm\i$ sets $t^{(1)}=0$.

For each scalar block, $\mathcal W^{(2)}$ contributes the mass term $\mathbf v_s^{(1)}\left(M_s(a_{i,s})\otimes I_{n_s}\right)\mathbf u_s^{(1)}$. Here $\mathbf u_s^{(1)}$ and $\mathbf v_s^{(1)}$ are the leading coefficients of $\Delta\mathbf u_s$ and $\Delta\mathbf v_s$, and $M_s(a_{i,s})$ is the background value of $M_s(A^{ss})$,
\begin{equation}\label{eq:first-uv-quadratic}
 M_s(a_{i,s})=\begin{pmatrix}
 -a_{1,s}&-1&t_{\rm core}\\
 1&-a_{2,s}&-1-G_s\\
 t_{\rm core}^{-1}&1+H_s&-a_{3,s}
 \end{pmatrix}.
\end{equation}
Note that the block adjoints $A_i^{ss,(1)}$ do not appear at this order. The core enters only through the background values of $G$ and $H$,
\begin{equation}\label{eq:first-uv-contractions}
\begin{aligned}
 G_s&=v_3^{\rm core}(A_1^{\rm core}-a_{1,s}I_R)^{-1}u_2^{\rm core},\\
 H_s&=v_2^{\rm core}(A_1^{\rm core}-a_{1,s}I_R)^{-1}u_3^{\rm core},
\end{aligned}
\end{equation}
which vanish for the empty core. The order $\delta$ equations for the rows and columes are
\begin{equation}\label{eq:first-uv-equations}
 \sum_j(M_s)_{ij}u_j^{s,(1)}=0,\qquad
 \sum_i v_i^{s,(1)}(M_s)_{ij}=0.
\end{equation}
The trivial solution $\mathbf u_s^{(1)}=\mathbf v_s^{(1)}=0$ exists for every position of the block, but the trace condition \eqref{eq:reduced-exact-CM-trace} excludes it. At first order, it requires $v_3^{s,(1)}u_3^{s,(1)}=-n_sc_3\neq0$. Nontrivial solutions exist precisely when $\det M_s(a_{i,s})=0$. Using the core identity $p_1q_1=z^2+1$ of \eqref{eq:core-pq-differential}, Appendix~\ref{app:second-order-blocks} shows that the core drops out of this determinant, which depends only on the scalar values in each block,
\begin{equation}\label{eq:second-scalar-surface}
 \det M_s(a_{i,s})=-\bigl(a_{1,s}a_{2,s}a_{3,s}+a_{1,s}-a_{2,s}+a_{3,s}\bigr).
\end{equation}
Therefore, the scalars in the moduli \eqref{eq:zero-source-moduli} are fixed to the cubic surface $\det M_s(a_{i,s})=0$ and the strings ending on each block $s$ develop a massless mode. This surface is smooth and $M_s(a_{i,s})$ has rank two on it, see Appendix~\ref{app:second-order-blocks}. Its right and left kernels are denoted by $\mathbf r_s$ and $\boldsymbol\ell_s$ given in \eqref{eq:second-kernel-components}, with the third components normalized to $1$. The massless mode is then given by $\mathbf u_s^{(1)}=\mathbf r_s u_{3}^{s,(1)}, \mathbf v_s^{(1)}=\boldsymbol\ell_s v_{3}^{s,(1)}$, while the other two directions are massive.

\subsubsection*{Order $\delta^2$ equations: six scalar sites and the Calogero-Moser space.}
Because the source term comes with the factor $\delta^2$, the zero-source moduli are first constrained at this order. We keep the $a_{i,s}$ off shell without yet imposing $\det M_s(a_{i,s})=0$. The massless mode identified above then gets a mass proportional to the determinant
\begin{equation}\label{eq:reduced-site-leading-W}
 \mathcal W^{(2)}\supset v_3^{s,(1)}\,u_3^{s,(1)}\,\frac{\det M_s(a_{i,s})}{a_{1,s}a_{2,s}+1}-n_s\sum_i c_i a_{i,s}.
\end{equation}
see \eqref{eq:second-kernel-components}-\eqref{eq:second-schur-bilinear}. The core part of $\mathcal W^{(2)}$ does not depend on the block scalar values.

In terms of $a_{i,s}$, the superpotential has the form of a linear potential subject to the constraint $\det M_s(a_{i,s})=0$, where the string condensate $v_3^{s,(1)}u_3^{s,(1)}$ plays the role of the Lagrange multiplier. Indeed, varying $u_3^{s,(1)}$ or $v_3^{s,(1)}$ imposes $\det M_s(a_{i,s})=0$. Varying $a_{i,s}$ on that surface gives
\begin{equation}\label{eq:reduced-site-stationarity}
 \frac{v_3^{s,(1)}u_3^{s,(1)}}{a_{1,s}a_{2,s}+1}\,
       \frac{\partial\det M_s}{\partial a_{i,s}}=n_sc_i.
\end{equation}
The third component is equivalent to the trace condition \eqref{eq:reduced-exact-CM-trace}. This is because varying $a_{i,s}$ corresponds to the shift $A_i^{ss}\mapsto A_i^{ss}+\epsilon_iI_{n_s}$, it is then equivalent to taking the trace of the corresponding adjoint equation.
Eliminating $v_3^{s,(1)} u_3^{s,(1)}$ from the other two components gives
\begin{equation}\label{eq:second-position-conditions}
 c_1=c_3\frac{a_{2,s}a_{3,s}+1}{a_{1,s}a_{2,s}+1},\qquad
 c_2=c_3\frac{a_{1,s}a_{3,s}-1}{a_{1,s}a_{2,s}+1}.
\end{equation}
Equivalently, $(c_1,c_2,c_3)$ is proportional to $-\nabla\det M_s(a_{i,s})=(a_{2,s}a_{3,s}+1,\,a_{1,s}a_{3,s}-1,\,a_{1,s}a_{2,s}+1)$. These critical conditions are independent of the core and the multiplicity $n_s$.

Eliminating $a_{2,s}$ and $a_{3,s}$ produces a sixth-order equation for $a_{1,s}$,
\begin{equation}\label{eq:second-six-positions}
\begin{gathered}
 (a_{1,s}^2+1)(c_1a_{1,s}^2+c_2-c_3)^2+4c_2c_3=0,\\
 a_{2,s}=\frac{c_1a_{1,s}^2+c_2+c_3}
                   {a_{1,s}(c_3-c_2-c_1a_{1,s}^2)},\qquad
 a_{3,s}=\frac{-c_1a_{1,s}^2+c_2+c_3}
                   {a_{1,s}(c_3-c_2-c_1a_{1,s}^2)}.
\end{gathered}
\end{equation}
It is cubic in $a_{1,s}^2$, with leading coefficient $c_1^2$ and value $(c_2+c_3)^2$ at $a_{1,s}=0$. For generic $c_i$ it has three distinct nonzero roots in $a_{1,s}^2$, which gives six distinct values of $a_{1,s}$. The denominators in \eqref{eq:second-six-positions} do not vanish at these roots: if $c_1a_{1,s}^2=c_3-c_2$, the polynomial reduces to $4c_2c_3\neq0$. Therefore, the critical points of the source within the cubic surface restricts the scalar positions in \eqref{eq:zero-source-moduli} to six triples. From now on, we label blocks by the six sites, $s=1,\ldots,6$, with occupation numbers $n_s\geq0$ and $\sum_sn_s=N-r(r+1)$.

If $a_{1,s}^2+1=0$, the polynomial reduces to $4c_2c_3\neq0$. The sites therefore avoid the core eigenvalues $\pm\i$. Their values of $a_{1,s}$ are also pairwise distinct, justifying our earlier assumption for the block gauge \eqref{eq:exact-block-gauge}.

The remaining order $\delta^2$ equations constrain the leading block adjoints. They follow by varying $\mathcal W^{(3)}$ with respect to the first-order coefficients. Once the massive directions are eliminated, the scalar block part of $\mathcal W^{(3)}$ is
\begin{equation}\label{eq:second-schur-cubic}
 \mathcal W^{(3)}\supset\sum_s\biggl[
 -\Tr_{n_s}A_1^{ss,(1)}[A_2^{ss,(1)},A_3^{ss,(1)}]
 +v_3^{s,(1)}\frac{\det M_s \left(A_{i}^{ss}\right)^{(1)}}{a_{1,s}a_{2,s}+1}u_3^{s,(1)}
 -\sum_ic_i\Tr_{n_s}A_i^{ss,(1)}\biggr].
\end{equation}
Its first and last terms are the cubic interaction and the linear source. The displacement of the $\det M_s=0$ surface by $A_{i}^{ss,(1)}$ supplies a mass term for the light modes in the middle. 
Expanding the mass matrix $M_s(A^{ss})$ to first order in $A_{i}^{ss,(1)}$, and contracting with the kernel vectors using Jacobi's formula \eqref{eq:resolvent-jacobi}, we find it involves only the normal combination
\begin{equation}\label{eq:second-normal-combination}
\begin{aligned}
 \beta_s^{(1)}\equiv-\det M_s \left(A_{i}^{ss}\right)^{(1)}={}&-\sum_i\frac{\partial \det M_s}{\partial a_{i,s}} A_i^{ss,(1)}\\
       ={}&(a_{2,s}a_{3,s}+1)A_1^{ss,(1)}
       +(a_{1,s}a_{3,s}-1)A_2^{ss,(1)}+(a_{1,s}a_{2,s}+1)A_3^{ss,(1)}\\
={}&\frac{a_{1,s}a_{2,s}+1}{c_3}
 \left(c_1A_1^{ss,(1)}+c_2A_2^{ss,(1)}+c_3A_3^{ss,(1)}\right).
\end{aligned}
\end{equation}

Varying the adjoints, the row, and the column and using \eqref{eq:second-position-conditions}, we find
\begin{subequations}\label{eq:second-light-system}
\begin{align}
 [A_2^{ss,(1)},A_3^{ss,(1)}]
 +\frac{c_1}{c_3}u_3^{s,(1)}v_3^{s,(1)}
 +c_1I_{n_s}&=0,\\
 [A_3^{ss,(1)},A_1^{ss,(1)}]
 +\frac{c_2}{c_3}u_3^{s,(1)}v_3^{s,(1)}+c_2I_{n_s}&=0,\\
 [A_1^{ss,(1)},A_2^{ss,(1)}]
 +u_3^{s,(1)}v_3^{s,(1)}+c_3I_{n_s}&=0,\\
 \beta_s^{(1)}u_3^{s,(1)}=0,\qquad
 v_3^{s,(1)}\beta_s^{(1)}&=0.
\end{align}
\end{subequations}
See derivations around \eqref{eq:order2fullequations}.
Its second and third adjoint equations are the leading terms of the exact constraints \eqref{eq:reduced-exact-CM}.
We now show that the last line of \eqref{eq:second-light-system} forces $\beta_s^{(1)}=0$. By \eqref{eq:second-normal-combination}, $\beta_s^{(1)}$ is proportional to $\sum_ic_iA_i^{ss,(1)}$. The three adjoint equations in \eqref{eq:second-light-system} then imply $[\beta_s^{(1)},A_2^{ss,(1)}]=[\beta_s^{(1)},A_3^{ss,(1)}]=0$, so the image of $\beta_s^{(1)}$ is invariant under $A_2^{ss,(1)}$ and $A_3^{ss,(1)}$. Since $v_3^{s,(1)}\beta_s^{(1)}=0$, the term $u_3^{s,(1)}v_3^{s,(1)}$ in the first adjoint equation vanishes on this image. Restricting that equation to the image and taking the trace gives $c_1\dim\operatorname{im}\beta_s^{(1)}=0$. Hence $\beta_s^{(1)}=0$ for $c_1\neq0$, which means $A_1^{ss,(1)}=-(c_2A_2^{ss,(1)}+c_3A_3^{ss,(1)})/c_1$. With this relation, the second and third adjoint equations in \eqref{eq:second-light-system} become $c_2/c_1$ and $c_3/c_1$ times the first. We are left with a single Calogero-Moser constraint
\begin{equation}\label{eq:second-CM-constraint}
 \Theta_s\equiv[A_2^{ss,(1)},A_3^{ss,(1)}]
   +\frac{c_1}{c_3}u_3^{s,(1)}v_3^{s,(1)}+c_1I_{n_s}=0.
\end{equation}

Later, we call $(A_2^{ss,(1)},A_3^{ss,(1)},u_3^{s,(1)},v_3^{s,(1)})$ the light fields of block $s$, since all other first-order fields are either determined by them or massive. Up to rescalings of the fields, \eqref{eq:second-CM-constraint} is the standard defining equation of a Calogero-Moser space~\cite{Wilson}, and its quotient by $\GL(n_s,\C)$ has complex dimension $2n_s$. In summary, order $\delta^2$ equations fix the positions of the scalar sites and leave Calogero-Moser spaces of total dimension $2n=2\sum_sn_s$, which are lifted at order $\delta^3$.

\subsubsection*{Order $\delta^3$ equations: quadratic potential and modified Calogero-Moser space.}
\label{sec:quadratic-potential}

The order $\delta^3$ equations $\mathcal E_\phi^{(3)}=\partial\mathcal W^{(4)}/\partial\phi^{(1)}=0$ follow from varying the $\delta^4$ coefficient $\mathcal W^{(4)}$. The third-order coefficients drop out of $\mathcal W^{(4)}$, since they are multiplied by the first-order equations $\mathcal E_\phi^{(1)}=0$ by \eqref{eq:W-F-order-counting}. The massive second-order coefficients, namely the core responses and the massive components of $u_i^{s,(2)}$ and $v_i^{s,(2)}$, are given in Appendix~\ref{app:second-order-blocks}. They can be solved for first and substituted into $\mathcal W^{(4)}$, since $\partial\mathcal W^{(4)}/\partial\phi^{(2)}=\mathcal E_\phi^{(2)}$ vanishes on the second-order solution. We are then left with a superpotential for the light fields and the second-order block adjoints $A_i^{ss,(2)}$ only. We first evaluate the terms of each block with the core at its background value, and then include the core responses and the couplings $W_{\rm links}$ between blocks.

Expanding \eqref{eq:second-schur-cubic} to one higher order gives
\begin{equation}\label{eq:third-schur-cubic}
\begin{aligned}
 \mathcal W^{(4)}\supset\sum_s\biggl[&
 -\sum_{\rm cyc}\Tr_{n_s}A_i^{ss,(2)}[A_j^{ss,(1)},A_k^{ss,(1)}]\\
 &+v_3^{s,(1)}\frac{\det M_s \left(A_{i}^{ss}\right)^{(2)}}{a_{1,s}a_{2,s}+1}u_3^{s,(1)}
 -\sum_ic_i\Tr_{n_s}A_i^{ss,(2)}\biggr].
\end{aligned}
\end{equation}
For the mass term in the middle, the kernel components of $u_i^{s,(2)}$ and $v_i^{s,(2)}$ drop out while the massive components contributes through \eqref{eq:second-uv-response}, see Appendix~\ref{app:quartic-reduction}. The second order displacements of the determinant supplies the mass for the light D2-D6 string, while order $\delta^0$ and $\delta$ contributions vanish by $\det M_s(a_{i,s})=0$ and $\beta_s^{(1)}=0$. With the core at its background value, $\det M_s$ is the cubic polynomial \eqref{eq:second-scalar-surface} in the $a_{i,s}$, which already includes the dependence of $G$ and $H$ on $A_1^{ss}$. Its value at the shifted positions can therefore be expanded as
\begin{equation}\label{eq:third-det-expansion}
\begin{aligned}
 &\det M_s\bigl(a_{i,s}I_{n_s}+\delta A_i^{ss,(1)}+\delta^2A_i^{ss,(2)}\bigr)\\
 &\quad=-\delta \beta_s^{(1)}
 -\delta^2\Bigl[\beta_s^{(2)}
   +\tfrac12\bigl(a_{3,s}\{A_1^{ss,(1)},A_2^{ss,(1)}\}\\
 &\hspace{25mm}+a_{2,s}\{A_1^{ss,(1)},A_3^{ss,(1)}\}
   +a_{1,s}\{A_2^{ss,(1)},A_3^{ss,(1)}\}\bigr)\Bigr]
 +O(\delta^3),
\end{aligned}
\end{equation}
where $\beta_s^{(2)}$ is \eqref{eq:second-normal-combination} with $A_i^{ss,(1)}$ replaced by $A_i^{ss,(2)}$. Substituting $A_1^{ss,(1)}=-(c_2A_2^{ss,(1)}+c_3A_3^{ss,(1)})/c_1$ and using \eqref{eq:second-position-conditions}, the mass term becomes
\begin{equation}\label{eq:third-schur-expansion}
v_3^{s,(1)}\Bigl(-\frac1{c_3}\,\sum_ic_iA_i^{ss,(2)}+q_s
   +\frac{c_1}{c_3}\,e_s\,[A_2^{ss,(1)},A_3^{ss,(1)}]\Bigr)u_3^{s,(1)},
\end{equation}
where $q_s$, the symmetric quadratic part of $\det M_s \left(A_{i}^{ss}\right)^{(2)}$, is given by
\begin{equation}\label{eq:third-quadratic}
\begin{gathered}
 q_s=q_{22,s}\bigl(A_2^{ss,(1)}\bigr)^2+q_{33,s}\bigl(A_3^{ss,(1)}\bigr)^2
 +\frac{q_{23,s}}2\bigl\{A_2^{ss,(1)},A_3^{ss,(1)}\bigr\},\\
 q_{22,s}=\frac{a_{3,s}(a_{1,s}a_{3,s}-1)}{(a_{1,s}a_{2,s}+1)(a_{2,s}a_{3,s}+1)},\qquad
 q_{23,s}=-\frac{2a_{1,s}}{(a_{1,s}a_{2,s}+1)(a_{2,s}a_{3,s}+1)},\\
 q_{33,s}=\frac{a_{2,s}}{a_{2,s}a_{3,s}+1}.
\end{gathered}
\end{equation}
Its coefficients depend only on the scalar values, and not on the core or on $n_s$. The last term in \eqref{eq:third-schur-expansion} is the antisymmetric part with $e_s=\kappa/(a_{2,s}a_{3,s}+1)^2$, where $\kappa$ is the core constant \eqref{eq:core-kappa} and $\kappa=t_{\rm core}^{-1}$ for the empty core, see Appendix~\ref{app:quartic-reduction}. We will see that $q_s$ gives the potential on the Calogero-Moser space, while the $e_s$ term only shifts the Lagrange multiplier so its value does not affect the counting.

We now collect the terms of \eqref{eq:third-schur-cubic} into a single trace for each block. Since $\partial\mathcal W^{(4)}/\partial\phi^{(2)}=\partial\mathcal W^{(3)}/\partial\phi^{(1)}$, $A_i^{ss,(2)}$ acts as Lagrange multiplier for the order $\delta^2$ adjoint equations \eqref{eq:second-light-system}. By $\beta_s^{(1)}=0$, all three are proportional to $\Theta_s$, therefore
$$
 -\frac1{c_1}\sum_ic_i\Tr_{n_s}\bigl(A_i^{ss,(2)}\Theta_s\bigr)
 =-\frac{c_3}{c_1(a_{1,s}a_{2,s}+1)}\Tr_{n_s}\bigl(\beta_s^{(2)}\Theta_s\bigr).
$$
Since $q_s$ is built from $(A_2^{ss,(1)})^2$, $(A_3^{ss,(1)})^2$ and their anticommutator, cyclicity of the trace gives $\Tr(q_s[A_2^{ss,(1)},A_3^{ss,(1)}])=0$. Hence
$$
 v_3^{s,(1)}q_su_3^{s,(1)}=\Tr_{n_s}\!\Bigl[\frac{c_3}{c_1}q_s\Theta_s-c_3q_s\Bigr].
$$
On the Calogero-Moser space this is the quadratic potential $-c_3\Tr_{n_s}q_s$.

We now include the remaining terms from the second-order core responses entering $W_{\rm core}$ and the dependence of $M_s(A^{ss})$ on $t$, $G$ and $H$, together with $W_{\rm links}$ quartic in the leading columns and rows. Both depend on the light fields only through the contractions $v_3^{s,(1)}u_3^{s,(1)}$, since $u_3^{s,(1)}$ and $v_3^{s,(1)}$ fix the other first-order columns and rows through the kernels of $M_s(a_{i,s})$. Together with the commutator term of \eqref{eq:third-schur-expansion}, these contributions are functions of the contractions up to terms proportional to $\Theta_s$. Since $\Tr\Theta_s=\frac{c_1}{c_3}v_3^{s,(1)}u_3^{s,(1)}+c_1n_s$, they give $\sum_s\Tr_{n_s}(\mathcal T_s\Theta_s)$ plus a constant, with $\mathcal T_s$ given in \eqref{eq:third-multiplier-shift}. Up to a constant, $\mathcal W^{(4)}$ reduces to
\begin{equation}\label{eq:third-effective-W}
\begin{gathered}
 \mathcal W_{\rm eff}^{(4)}=\sum_s\Tr_{n_s}\!\left[\Lambda_s\Theta_s-c_3q_s\right],\\
 \Lambda_s=-A_1^{ss,(2)}-\frac{c_2}{c_1}A_2^{ss,(2)}-\frac{c_3}{c_1}A_3^{ss,(2)}
   +\frac{c_3}{c_1}q_s+\mathcal T_s.
\end{gathered}
\end{equation}
Since second order adjoints $A_i^{ss,(2)}$ appears only in $\Lambda_s$, we can use $\Lambda_s$ as an independent variable. It becomes the Lagrange multiplier for the Calogero-Moser constraint, and the critical points of \eqref{eq:third-effective-W} are the stationary points of quadratic potential $\Tr q_s$ on the Calogero-Moser space.

To find these critical points, we bring \eqref{eq:third-effective-W} to the conventions of \cite{Felder:2024thr}. For generic sources the quadratic form $q_s$ is nondegenerate, and we choose a nonzero $\omega_s$ with $\omega_s^2=q_{23,s}^2-4q_{22,s}q_{33,s}$. Under the linear map
\begin{equation}\label{eq:third-paper-map}
\begin{gathered}
 \mathsf L_s=X'_s+\frac{Y'_s}{2c_1},\qquad
 \mathsf Q_s=X'_s-\frac{Y'_s}{2c_1},\\
 \mathsf M_s=\frac{\Lambda_s}{c_3\omega_s},\qquad
 \mathsf v_s=-\frac{u_3^{s,(1)}}{c_3},\qquad \mathsf w_s=v_3^{s,(1)},\\
 X'_s=A_2^{ss,(1)}+\frac{q_{23,s}-\omega_s}{2q_{22,s}}A_3^{ss,(1)},\qquad
 Y'_s=\frac{q_{22,s}}{\omega_s}A_2^{ss,(1)}
       +\frac{q_{23,s}+\omega_s}{2\omega_s}A_3^{ss,(1)},
\end{gathered}
\end{equation}
the effective superpotential \eqref{eq:third-effective-W} becomes
$$
 \mathcal W_{\rm eff}^{(4)}
 =\sum_s c_1c_3\omega_s\Tr_{n_s}\!\left[
   \mathsf M_s\bigl(I_{n_s}-\mathsf v_s\mathsf w_s-[\mathsf L_s,\mathsf Q_s]\bigr)
   -\frac12\bigl(\mathsf L_s^2-\mathsf Q_s^2\bigr)\right].
$$

Varying $\mathsf M_s$, $\mathsf L_s$, $\mathsf Q_s$, $\mathsf v_s$ and $\mathsf w_s$ gives
\begin{equation}\label{eq:third-modified-CM}
\begin{gathered}
 [\mathsf L_s,\mathsf Q_s]=I_{n_s}-\mathsf v_s\mathsf w_s,\qquad
 [\mathsf M_s,\mathsf Q_s]=\mathsf L_s,\qquad
 [\mathsf M_s,\mathsf L_s]=\mathsf Q_s,\\
 \mathsf M_s\mathsf v_s=0,\qquad \mathsf w_s\mathsf M_s=0,
\end{gathered}
\end{equation}
which are the defining relations of the modified Calogero-Moser space of \cite{Felder:2024thr}.

In terms of $\mathsf Z_{\pm}=\mathsf L_s\pm\mathsf Q_s$, the last four relations read $[\mathsf M_s,\mathsf Z_\pm]=\pm\mathsf Z_\pm$, $\mathsf M_s\mathsf v_s=0$ and $\mathsf w_s\mathsf M_s=0$. Up to the gauge transformation generated by $\mathsf M_s$, a solution is therefore invariant under the $\C^\ast$ action $\mathsf Z_\pm\mapsto\lambda^{\pm1}\mathsf Z_\pm$ on the Calogero-Moser space, and $\Tr(\mathsf L_s^2-\mathsf Q_s^2)=\Tr\mathsf Z_+\mathsf Z_-$ is the Hamiltonian that generates this action. The vacua of block $s$ are thus the fixed points of this $\C^\ast$ action, which are classified by Wilson~\cite{Wilson}.

\subsubsection*{Solutions labelled by partitions.}
Theorem~3 of \cite{Felder:2024thr} assigns exactly one solution of \eqref{eq:third-modified-CM} to each partition of $n_s$, $\nu_s=(\nu_{s,1}\ge\cdots\ge\nu_{s,\ell}>0)$ with $\sum_a\nu_{s,a}=n_s$. Its Young diagram has $\nu_{s,a}$ boxes in row $a$. Write $\operatorname{Spec}$ for the set of eigenvalues counted with multiplicity. Propositions~6 and~8 of \cite{Felder:2024thr} give the spectra of the solution matrices. In particular, the spectrum of the multiplier is
\begin{equation}\label{eq:third-partition-spectrum}
 \operatorname{Spec}\mathsf M_s=\{j-i:(i,j)\text{ is a box of }\nu_s\},
\end{equation}
where $i$ and $j$ are the row and column numbers of a box starting at $1$. Appendix~\ref{app:partition-solutions} collects the spectra of $\mathsf Q_s$ and $\mathsf L_s$, explicit representatives of each gauge orbit and provides an example for $n_s=3$.

Each partition is a nondegenerate critical point. By Section~4 and Appendix~A.3 \cite{Felder:2024thr}, the tangent weights of the $\C^\ast$ action are the pairs $\pm h(\Box)$ over the boxes of $\nu_s$, where the hook length $h(\Box)$ counts the box and the boxes to its right and below it. Since the weight are nonzero, the Hessian of the constrained potential is invertible.
One can then continue the perturbation calculations based on these light mode solutions. The equations for the massive and normal directions can be solved analytically in terms of the light fields. The higher order coefficients of the light fields are solved near the critical partition points using the invertible Hessian.

For a fixed core and occupation numbers $(n_1,\ldots,n_6)$, the perturbation theory therefore gives
\begin{equation}\label{eq:third-partition-count}
 \prod_{s=1}^{6}p(n_s),\qquad
 \sum_{s=1}^{6}n_s=N-r(r+1),\qquad p(0)=1,
\end{equation}
local branches. Section~\ref{sec:numerical} tests these structures using exact solutions and the accuracy of the perturbative series. Section~\ref{sec:SummaryofCounting} sums these multiplicities over occupations and core sectors to obtain the BPS count.

\subsection{Numerical checks}\label{sec:numerical}

\begin{figure}[t]
 \centering
 \includegraphics[width=\textwidth]{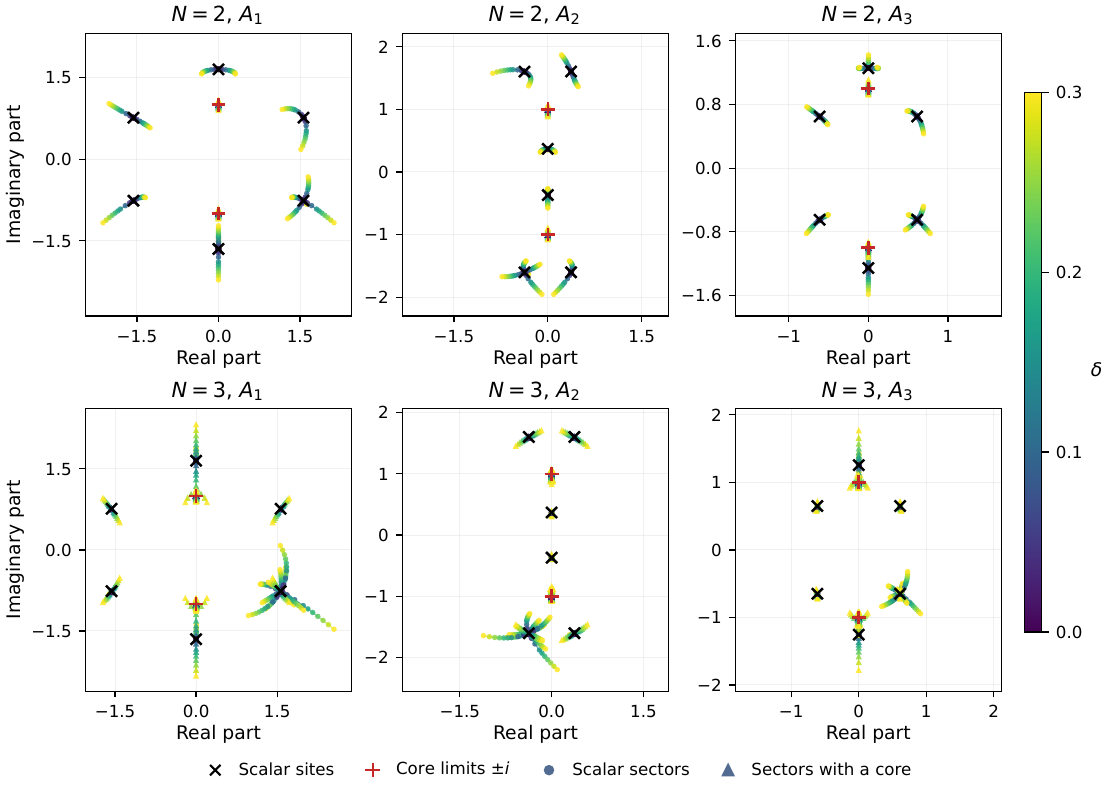}
 \caption{Finite-source spectra of $A_1,A_2,A_3$ at \eqref{eq:numerical-parameters} for $N=2$ (top row) and $N=3$ (bottom row). All branches are followed over $10^{-8}\le\delta\le0.3$, with colour denoting $\delta$. Black crosses mark the six scalar sites and red plus signs the core limits $\pm\i$. All axes are linear.}
 \label{fig:numerical-spectra}
\end{figure}

We now test the construction numerically. For $N=2$ and $N=3$ we solve the full F-equations without expanding in $\delta$ using the method in \cite{Chowdhury:2015gbk}. We take the parameters
\begin{equation}\label{eq:numerical-parameters}
 (c_1,c_2,c_3)=(1,2,3),\qquad c^{(i4)}=c_i\delta^2,
\end{equation}
which satisfy the genericity conditions of Section~\ref{sec:NonzeroSource}, and $10^{-8}\le\delta\le0.3$. We follow ten representative branches for $N=2$ and seventeen for $N=3$. They include all purely scalar sectors and the two $r=1$ cores. All roots satisfy the F-equations with residuals below $10^{-40}$.
Figure~\ref{fig:numerical-spectra} shows the eigenvalues of $A_1$, $A_2$ and $A_3$, which are gauge invariant. As $\delta\to0$, the $r=0$ spectra approach the six scalar sites \eqref{eq:second-six-positions}, with multiplicities given by the occupation numbers. An $r=1$ core instead contributes $+\i,-\i$ to the spectrum, together with one additional scalar eigenvalue for $N=3$. The core and scalar sectors are thus clearly separated. 

\begin{figure}[t]
 \centering
 \includegraphics[width=\textwidth]{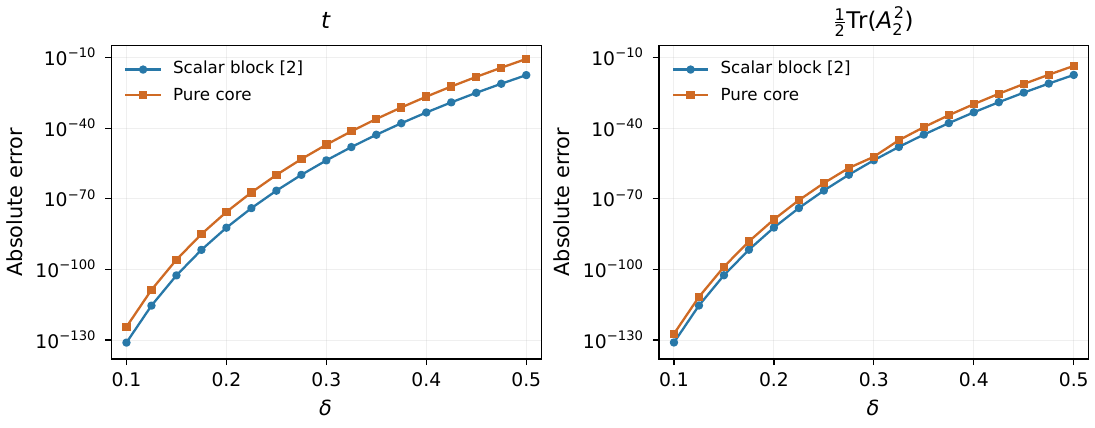}
 \caption{Absolute errors of the order 160 field expansion at $N=2$. The scalar branch has partition $(2)$ at the site with $a_{2,s}\simeq-0.372059-1.603210\,\i$ while the pure-core branch has $r=1$. Both have $t(0)=+\i$. The observables are evaluated on the partial sums and compared with solutions of the untruncated F-equations.}
 \label{fig:numerical-order160}
\end{figure}

For $N=2$ we also solve the perturbative equations of Section~\ref{sec:small-delta} through order $160$. This allows a direct comparison between the series sum $\phi^{[160]}(\delta)=\sum_{m=0}^{160}\delta^m\phi^{(m)}$ and the exact solutions $\phi_{\rm exact}(\delta)$ of the full F-equations. Figure~\ref{fig:numerical-order160} shows the absolute errors $|\mathcal O(\phi^{[160]})-\mathcal O(\phi_{\rm exact})|$ for two gauge-invariant observables, $\mathcal O=t$ and $\tfrac12\Tr(A_2^2)$. They are shown on a scalar branch with partition $(2)$ and on a pure-core branch, over $0.1\le\delta\le0.5$. The largest error among these observables and sampled points is $2.7\times10^{-11}$ attained by $t$ on the core branch at $\delta=0.5$. The perturbative construction therefore agrees with the solutions of the exact F-equations up to $\delta=0.5$. This agreement indicates that the perturbative construction persist at finite $\delta$.

\section{Putting everything together}
\label{sec:SummaryofCounting}

We now combine the results of the previous section to obtain an explicit count of isolated BPS configurations at finite $c^{(i4)}$ and compare this count with predictions from string dualities and the black hole solution.

The BPS configurations are labeled in part by a nonnegative integer $r$, which fixes the core dimension to $R=r(r+1)$, see \eqref{eq:core-dimension}. At fixed $N$, the allowed values satisfy $r(r+1)\leq N$. For each $r$, there are two rigid core solutions at $c^{(i4)}=0$, which remain rigid under small deformations to nonzero $c^{(i4)}$, see \eqref{eq:core-leading-order} and Appendix~\ref{app:core-mass}. Once $r$ is fixed, the remaining $n=N-r(r+1)$ D6 directions are described at $c^{(i4)}=0$ by scalar blocks with moduli space $(\C^3)^n/S_n$ in \eqref{eq:zero-source-moduli}, the eigenvalues of $A_i$ along those blocks. Turning on generic nonzero sources selects six possible limiting scalar triplets, given by \eqref{eq:second-six-positions}, which we denoted by $a_{i,s}$ with $s=1,\ldots,6$. Their occupancies are specified by 
\beq
(n_1,\ldots,n_6) \in \mathbb{Z}^6_{\geq 0},\qquad \sum_{s=1}^6 n_s=n.
\eeq
where $n_s$ counts the D6 directions with limiting eigenvalue triple $(a_{1,s},a_{2,s},a_{3,s})$ as $c^{(i4)}=0$. These multiplicities do not fully specify a vacuum. When $c^{(i4)}$ is turned on, each block of dimension $n_s$ admits $p(n_s)$ isolated deformations, one for each partition of $n_s$, see \eqref{eq:third-modified-CM} and \eqref{eq:third-partition-spectrum}. Taking into account that there are two isolated cores (related by complex conjugation) the number of BPS configurations is, by \eqref{eq:third-partition-count},
\beq
(n_1,\ldots,n_6)\qquad \Rightarrow \qquad 2\, p(n_1)\ldots p(n_6),
\eeq
whenever the integers satisfy the sum rule that the total dimension is $n=N-r(r+1)$ and otherwise is zero.

The total number of BPS states is therefore given by a sum over $r$, and for each $r$ we have the product of partitions of sixtuplets that add to $N-r(r+1)$. In other words,
\begingroup
\setlength{\abovedisplayskip}{15pt}
\setlength{\belowdisplayskip}{15pt}
\setlength{\abovedisplayshortskip}{15pt}
\setlength{\belowdisplayshortskip}{15pt}
\beq
d_N
=\sum_{\substack{r\ge0\\r(r+1)\le N}}
\sum_{\substack{n_1,\ldots,n_6\geq0\\
n_1+\cdots+n_6=N-r(r+1)}}
2\,p(n_1)\ldots p(n_6).
\label{eq:index-N}
\eeq
\endgroup
This is the index of the D2/D2/D2/D6 quantum mechanics for a stack of $N$ D6-branes. The main result of this paper is having proven this, starting from the D-brane quantum mechanics and counting BPS vacua. From this general formula we can recover some particular examples from \cite{Chowdhury:2014yca,Chowdhury:2015gbk}. As we mentioned in Section~\ref{sec:ZeroSourceSolutions}, $d_0=2$ since we only have the scalar fields at $t=\pm \i$. In the presence of D6 branes we have
\begin{itemize}
\item $N=1$: Only $r=0$ contributes since $r=1$ already requires $2\times 2$ matrices. The single scalar direction can occupy any of the six positions, giving
\beq
d_1=2\times6=12.
\eeq

\item $N=2$: For $r=0$, the two scalar directions either occupy the same position, contributing $6p(2)$, or two distinct positions, contributing $\binom{6}{2}$. The $r=1$ core uses all two D6 directions and contributes one configuration per branch. Thus
\beq
d_2=2\left[6\,p(2)+\binom{6}{2}+1\right]
=2(12+15+1)=56.
\eeq

\item $N=3$: For $r=0$, the possible occupancy patterns are $3$, $2+1$, and $1+1+1$, contributing $6p(3)$, $6\cdot5\cdot p(2)$, and $\binom{6}{3}$, respectively. For $r=1$, one scalar direction remains and can occupy any of the six positions. Therefore
\beq
\begin{aligned}
d_3&=2\left[6p(3)+6\cdot 5 \cdot p(2)+\binom{6}{3}+6\right]\
&=2(18+60+20+6)=208.
\end{aligned}
\eeq
\end{itemize}

This match persists for higher $N$ reproducing also the results of a more sophisticated numerical evaluation \cite{Chowdhury:2023wss,Chowdhury:2026noq}. Can show directly that for all $N$, our result for the counting matches the prediction from string theory we presented in the introduction? For this compute the generating function 
\beq
Z(q) = \sum_{N\geq 0} d_N \,q^{N},
\eeq
of the $N$-dependent index in \eqref{eq:index-N}. First of all, it is well-known that the generating function for partitions of integers is
$
\sum_{n\geq0}p(n)q^n=\prod_{k\geq1}(1-q^k)^{-1}.
$
The generating function of $d_N$ can be easily computed using this elementary result. The reason is that once we sum over $N$ the restriction over $r$ and therefore the sum $n_1+ \ldots + n_6$ goes away. The result is simply
\beq
Z(q)
 =2\sum_{r=0}^{\infty}\frac{q^{r(r+1)}}{\prod_{k\geq1}(1-q^k)^6}.
  \label{eq:count-product-series}
\eeq
Applying a Fourier transform of the right hand side as $q \to 1$ we can extract the large $N$ behavior of the index
\begin{equation}
 \log d_N=2\pi\sqrt N+O(\log N).
 \label{eq:gravity-entropy}
\end{equation}
This result matches the index of the Reissner-Nordstrom black hole dual to the D-brane quantum mechanics in the large charge limit.

We can use this generating function to show that our degeneracies are given by $-\widehat{c}(4N)$ as defined implicitly by equation \eqref{eq:thetas}. First notice that we can obtain a generating function for $\widehat{c}(4N)$ by
\beq
\sum_{n\geq0}
\widehat c(4n)\,q^n = - \frac{1}{\eta(\tau)^6} \oint \frac{\d y }{y} \, \vartheta_1(z|\tau)^2,
\label{eq:c-gen-first}
\eeq
where the integral on the RHS is done over a closed contour around the origin and its role is just to pick the coefficient $y^0$ in the expansion. To evaluate the RHS we can use the definition of the theta function and write the integrand as a sum over two indices
\beq
-\vartheta_1(z|\tau)^2
=
\sum_{a,b\in\mathbb Z}
(-1)^{a+b}
q^{\frac12[(a+1/2)^2+(b+1/2)^2]}
y^{a+b+1}.
\eeq
The integral over $y$ selects $b+a=-1$ leading to 
\beq
-\oint \frac{\d y }{y} \, \vartheta_1(z|\tau)^2 = - q^{1/4} \sum_{a\in\mathbb{Z}} q^{a(a+1)}=-2 q^{1/4} \sum_{r\geq 0} q^{r(r+1)},
\eeq
where on the second equality we restricted to $r>0$ by pairing $a=r$ with $a=-r-1$. Combining this with \eqref{eq:c-gen-first} gives
\beq
\sum_{n\geq0}
\widehat c(4n)\,q^n = - Z(q),\qquad \Rightarrow \qquad d_N = - \widehat c(4N).
\eeq
Therefore our analytic counting is a highly nontrivial check of the results in \cite{Maldacena:1999bp,Shih:2005qf,Pioline:2005vi,Sen:2008ta}. The same coefficients follow from the gravitational path integral
by supersymmetric localization~\cite{Iliesiu:2022kny}. This also automatically implies that the number of BPS configurations of the D-brane quantum mechanics matches the classical Bekenstein-Hawking extremal entropy at large $N$.

\subsection*{Bootstrapping the D-brane action} 

We conclude this analysis by considering a question raised by Ashoke Sen. Is the superpotential of the quiver quantum mechanics derived from string theory special, or would any theory with the same interaction structure reproduce the same index? Once the quiver structure is fixed, the interactions are largely determined. Up to field rescalings, the main remaining freedom can be captured by an overall rescaling of
\beq
W_{ZZZ} \to c \, W_{ZZZ},
\eeq
with arbitrary $c>0$. Does the BPS counting exhibit any pathology for $c\neq 1$, or could we equally well choose any positive value of $c$? We emphasize that $c=1$ is fixed by a worldsheet string theory calculation in \cite{Chowdhury:2015gbk}.

To answer this question, we compute the index as a function of $c$. Most of the derivation in section \ref{sec:counting} carries over unchanged. The main modification concerns the properties of the core, particularly the equation for $t$, or equivalently for the adjoint eigenvalues $a$. Repeating the derivation of the polynomials $p(z)$ and $q(z)$, which determine the Jordan form of the adjoints, yields
\beq
p(z)=\frac{(z-a)^{r+1}}{(z-a^{-1})^r},
\qquad
q(z)=\frac{(z-a^{-1})^{r+1}}{(z-a)^r},\qquad t= a^{-2r-1}
\eeq
where the eigenvalues $a^{\pm 1}$ satisfy the equation
\beq
(c^{-2}-1)(a^{4r+2}+1)+a^{2r+2}+a^{2r}=0.
\eeq
For $c\neq 1$, there are $4r+2$ solutions, corresponding to $4r+2$ isolated cores. At $c=1$, the polynomial degenerates, leaving only two nontrivial solutions. Thus, for any $c\neq 1$, the count is
\beq
d_N(c)=
\sum_{\substack{r\ge0\\r(r+1)\le N}}
\sum_{\substack{n_1,\ldots,n_6\ge0\\
n_1+\cdots+n_6=N-r(r+1)}}
(4r+2)\,p(n_1)\ldots p(n_6).
\eeq
For small $N$, it precisely reproduces the BPS counts obtained in \cite{Chowdhury:2015gbk} using the incorrect potential. This counting is also encoded in the generating function obtained by extracting the $y^1$ coefficient of the Jacobi form $- 2q^{1/4}/\vartheta^2_1(\tau|z)$. This result may already seem suspicious in light of the connection to the D1/D5/KK system. 

We now return to the original question by examining the semiclassical large $N$ limit of the index for general $c$. We find
\beq
\log d_N(c)=
\begin{cases}
2\pi\sqrt{N}-\dfrac{7}{4}\log N+O(1),
& c\neq1,\\[6pt]
2\pi\sqrt{N}-2\log N+O(1),
& c=1,
\end{cases}
\qquad N\to\infty.
\eeq
Both indices reproduce the classical Bekenstein-Hawking entropy, which might suggest that they are equally valid. Their one-loop corrections, however, differ. An explicit gravitational path integral calculation gives the logarithmic correction $-2 \log N$ \cite{Banerjee:2011jp}. Thus, even without invoking the duality to the D1/D5/KK system, the string theory value $c=1$ is singled out by consistency with gravity! Conversely, matching the large-$N$ behavior of the index to the gravitational path integral fixes $c$ and hence the corresponding worldsheet amplitude.

\section{Conclusions and future directions}\label{sec:conclusion}

We conclude by considering whether the D-brane quantum mechanics described in Section~\ref{sec:qm-review} provides a microscopic description of a near-extremal black hole in asymptotically flat spacetime. The attractor geometry is $AdS_2 \times S^2 \times T^6$, and a straightforward calculation gives the relevant parameters
\beq
\frac{L_{AdS_2}^2}{L_{Planck}^2} =2 \sqrt{N_1 N_2 N_3 N_4} ,\qquad \text{Vol}(T^6) = (2\pi)^6 \sqrt{\frac{N_1 N_2 N_3 }{N_4^3}},\qquad g_2 = e^{\phi} \left( \frac{N_1 N_2 N_3}{N_4^3}\right)^{1/4},
\label{eq:attractor}
\eeq
where $\phi$ is the 4d dilaton. For $N_1=N_2=N_3=1$ and $N_4=N$, the AdS radius in 4d Planck units is given by $2 \sqrt{N} \gg 1$. In the large $N$ limit, the string coupling vanishes as $g_s \sim 1/N^{3/4}$. However, the volume of the compact six-torus also shrinks in string units, $\text{Vol}(T^6) \sim 1/N^{3/2}$. This gives rise to a tower of light winding string modes that may invalidate the gravitational description. We return to this regime later in this section.

We can avoid this complication by considering multiple D2-branes wrapping each $T^2 \subset T^6$. The corresponding generalization of the superpotential in section \ref{sec:qm-review} is given in Appendix A of \cite{Chowdhury:2015gbk}. The multiplets associated with the D2-branes are now also non-abelian, while the structure of the interaction terms remains the same. We consider a limit in which all four charges scale as $(N_1,\ldots,N_4)=N(n_1,\ldots,n_4)$, with the $n_i$ held fixed as $N\to\infty$. The action then takes the form
\beq
S = \frac{N}{\lambda} \int \d t  \,\,\text{Tr} \Big[\, \text{kinetic} - V_F - V_D - V_{\text{gauge}} + \text{fermions}\,\Big].
\eeq
We denote by $\lambda$ the overall dimensionful 't Hooft coupling evaluated at the string scale, the ratio of the open string coupling on the $i$th D2-brane stack to that on the D6-brane stack is $\sqrt{n_i/n_4}$. The resulting theory has the structure expected for a standard 't Hooft limit. Using \eqref{eq:attractor}, the extremal entropy of the dual black hole is $S = 2 \pi N^2 \sqrt{n_1\ldots n_4}$. The size of the compact space remains finite in string units because the overall factor of $N$ cancels from the ratio in \eqref{eq:attractor}. The horizon size in Planck units is controlled by $N$, while its size in string units is controlled by $\lambda$. We work in the regime $1 \ll \lambda \ll N$, with $\lambda$ expressed in string units. Since the coupling is dimensionful, the effective coupling grows at low energies, and the system is strongly coupled at energies or temperatures well below the string scale. The 1d conformal symmetries of the throat may emerge in this regime.

Testing this proposal requires studying observables beyond the protected BPS index, which serves as a consistency check on the proposed gravity dual. Possible numerical tests include:

\begin{itemize}
    \item A useful test would be to perform Monte Carlo simulations along
the lines of~\cite{Anagnostopoulos:2007fw} and determine the thermal
energy above the BPS bound, after removing the free collective
degrees of freedom. In the semiclassical near-extremal regime,
the nearly-$AdS_2\times S^2\times T^6$ throat predicts a leading
correction quadratic in temperature, with a coefficient fixed by
the black hole solution. At lower temperatures, quantum effects
become important: the $\mathcal N=4$ super-Schwarzian
theory~\cite{Heydeman:2020hhw} predicts a definite relation between
this thermodynamic coefficient and the gap separating the BPS
states from the lowest non-BPS black hole excitations. Measuring
both quantities would provide a quantitative test of the proposed
gravitational description beyond the protected BPS counting.
    
    \item The semiclassical analysis of the F- and D-term equations determines
the BPS index, but does not determine the quantum wavefunctions of
the BPS states. Computing these wavefunctions would allow us to
apply the diagnostics of BPS chaos proposed
in~\cite{Lin:2022zxd,Chen:2024oqv}, probing information about the
BPS sector beyond its degeneracy. Another interesting direction
is to formulate a notion of fortuity appropriate to this family
of black holes, which is being pursued in~\cite{wip}. It would also be interesting to see whether the non-BPS states present a statistics similar to the random matrix of \cite{Turiaci:2023jfa}.

    \item Following the proposal of~\cite{Biggs:2023mfn}, numerical calculations
of real-time thermal correlation functions would provide a further
test of the gravitational description. In particular, their
relaxation at large $N$ could be compared with the low-energy
quasinormal frequencies associated with the throat. This offers a particularly tractable test, since the
corresponding near-horizon gravity calculation is considerably
simpler than in BFSS.
\end{itemize}

Let us return to the $(1,1,1,N)$ system. T-duality along all six torus directions maps its attractor region to a D4–D4–D4–D0 configuration whose type IIA coupling grows as $N^{3/4}$. Its M-theory lift consists of three wrapped M5-branes sharing the M-theory circle and carrying $N$ units of momentum along it\footnote{We thank Ashoke Sen for suggesting this.}. This motivates an infrared description in terms of the $(0,4)$ MSW conformal field theory. The large-momentum entropy $S\sim2\pi\sqrt N$ suggests that the central charge is $c=6$. This limit increases the momentum while keeping the central charge finite. The modular bootstrap suggests this limit is universal and described by an emergent broken 1d conformal symmetry \cite{Benjamin:2019stq,Ghosh:2019rcj, Heydeman:2020hhw}. It would be interesting to determine numerically whether the near-BPS sector actually has these features.

Finally, as discussed in the introduction, it would be valuable to extend the index calculation to more general systems. One direction is to allow arbitrary numbers of D2-branes, a problem currently under investigation in \cite{wip}. Another is to consider systems with less supersymmetry. A natural starting point is to reproduce the counting for $\mathcal{N}=4$ supergravity and then for $\mathcal{N}=2$ such as the D4-D0 system on the quintic studied for example in~\cite{Gaiotto:2005rp}.

\paragraph{Acknowledgments} 
We thank Chi-Ming Chang, Sarthak Duary, Haoyu Zhang, for useful discussions and collaboration on related matters. We are grateful to Ashoke Sen for discussions and comments on a draft. We used an internal Google research agent powered by Gemini Deep Think and the ChatGPT for Academic Researchers. AI was used to identify and process the mathematical literature with the theorems we needed, related to quiver varieties  and geometric invariant theory \cite{King94,Hoskins2014,CB01,Przezdziecki:2016} and on Calogero-Moser spaces \cite{Wilson,Felder:2024thr}.

\begin{appendix}

\section{Reduced superpotential and small $\delta$ expansion}
\label{app:ExpansionofW}

This appendix collects the equations and derivations used in Section~\ref{sec:NonzeroSource}. Appendix~\ref{app:exact-W} lists the terms of \eqref{app:zero:expansion}, derives the reduced superpotential and gives its F-equations. Appendix~\ref{app:second-order-blocks} contains the order $\delta$ and $\delta^2$ equations, and Appendix~\ref{app:quartic-reduction} collects the order $\delta^3$ computation.

\subsection{Exact expansion and reduction of the superpotential}
\label{app:exact-W}
We list the terms of \eqref{app:zero:expansion}. The core contribution to the quadratic term is
\begin{equation}\label{app:zero:core-quadratic}
\begin{aligned}
 \mathcal Q_{\rm core}={}&
 -\sum_{\rm cyc}\Tr_R A_i^{\rm core}
       [\Delta A_j^{\mathrm{cc}},\Delta A_k^{\mathrm{cc}}]\\
 &-\sum_i\bigl(
 \Delta v_i^{\mathrm c}A_i^{\rm core}\Delta u_i^{\mathrm c}
 +v_i^{\rm core}\Delta A_i^{\mathrm{cc}}\Delta u_i^{\mathrm c}
 +\Delta v_i^{\mathrm c}\Delta A_i^{\mathrm{cc}}u_i^{\rm core}\bigr)\\
 &+\sum_{i,j}K_{ij}(t_{\rm core})
                 \Delta v_i^{\mathrm c}\Delta u_j^{\mathrm c}\\
 &+\Delta t\biggl[
 \Delta v_1^{\mathrm c}u_3^{\rm core}
       +v_1^{\rm core}\Delta u_3^{\mathrm c}
 -\frac{1}{t_{\rm core}^2}
   \bigl(\Delta v_3^{\mathrm c}u_1^{\rm core}
                 +v_3^{\rm core}\Delta u_1^{\mathrm c}\bigr)\biggr]\\
 &+\frac{v_3^{\rm core}u_1^{\rm core}-1}{t_{\rm core}^3}
       (\Delta t)^2,
\end{aligned}
\end{equation}
where $K(t)$ is defined in \eqref{eq:Kdefinition} and cyclic sums run over the three cyclic permutations $(i,j,k)$ of $(1,2,3)$. For each scalar block $s$, the columns and rows and the core-block adjoints contribute
\begin{equation}\label{app:zero:core-site-quadratic}
\begin{aligned}
 \mathcal Q_{\mathrm cs}={}&
 \sum_{\rm cyc}\Tr_{n_s}\!\Bigl[
 \Delta A_i^{s\mathrm c}\bigl(
 (A_k^{\rm core}-a_{k,s}I_R)\Delta A_j^{\mathrm cs}\\
 &\hspace{32mm}
 -(A_j^{\rm core}-a_{j,s}I_R)\Delta A_k^{\mathrm cs}\bigr)\Bigr]\\
 &-\sum_i\bigl(v_i^{\rm core}\Delta A_i^{\mathrm cs}\Delta u_i^s
       +\Delta v_i^s\Delta A_i^{s\mathrm c}u_i^{\rm core}\bigr)\\
 &+\sum_{i,j}K_{ij}(t_{\rm core})\Delta v_i^s\Delta u_j^s
       -\sum_i a_{i,s}\Delta v_i^s\Delta u_i^s.
\end{aligned}
\end{equation}
The terms joining distinct scalar blocks are
\begin{equation}\label{app:zero:site-site-quadratic}
 \mathcal Q_{\ell m}=
 \sum_{\rm cyc}\Tr_{n_m}\!\left[\Delta A_i^{m\ell}
 \bigl((a_{k,\ell}-a_{k,m})\Delta A_j^{\ell m}
       -(a_{j,\ell}-a_{j,m})\Delta A_k^{\ell m}\bigr)\right].
\end{equation}
The cubic and higher degree interaction terms are
\begin{equation}\label{app:zero:remainder}
\begin{aligned}
 \mathcal R(\Delta\phi)={}&
 -\Tr\Delta A_1[\Delta A_2,\Delta A_3]
 -\sum_i\Delta v_i\Delta A_i\Delta u_i\\
 &+\Delta t\Delta v_1\Delta u_3
 -\frac{\Delta t}{t_{\rm core}(t_{\rm core}+\Delta t)}
                    \Delta v_3\Delta u_1\\
 &+\frac{(\Delta t)^2}{t_{\rm core}^2(t_{\rm core}+\Delta t)}
       \bigl(v_3^{\rm core}\Delta u_1^{\mathrm c}
                  +\Delta v_3^{\mathrm c}u_1^{\rm core}\bigr)\\
 &+\frac{(1-v_3^{\rm core}u_1^{\rm core})(\Delta t)^3}
            {t_{\rm core}^3(t_{\rm core}+\Delta t)}.
\end{aligned}
\end{equation}
Each term of $\mathcal R$ starts at degree three, and the terms of higher degree come from expanding $1/(t_{\rm core}+\Delta t)$, which converges for $|\Delta t|<|t_{\rm core}|$.

Differentiating \eqref{app:zero:expansion} with respect to the flutuation  gives the F-equations
\begin{align}
0={}&[A_j^{(0)},\Delta A_k]+[\Delta A_j,A_k^{(0)}]
       +[\Delta A_j,\Delta A_k]\notag\\
 &+u_i^{(0)}\Delta v_i+\Delta u_i v_i^{(0)}
       +\Delta u_i\Delta v_i+\delta^2c_iI_N,
       \qquad(i,j,k)\ \text{cyclic},\label{eq:exact-fluctuation-adj}
\\
0={}&A_i^{(0)}\Delta u_i+\Delta A_i(u_i^{(0)}+\Delta u_i)
       -\sum_jK_{ij}(t)\Delta u_j\notag\\
 &-\sum_j\bigl[K_{ij}(t)-K_{ij}(t_{\rm core})\bigr]u_j^{(0)},\label{eq:exact-fluctuation-column}
\\
0={}&\Delta v_iA_i^{(0)}+(v_i^{(0)}+\Delta v_i)\Delta A_i
       -\sum_j\Delta v_jK_{ji}(t)\notag\\
 &-\sum_jv_j^{(0)}\bigl[K_{ji}(t)-K_{ji}(t_{\rm core})\bigr]\label{eq:exact-fluctuation-row},
\\
0={}&v_1^{(0)}\Delta u_3+\Delta v_1u_3^{(0)}
          +\Delta v_1\Delta u_3\notag\\
 &-\frac{v_3^{(0)}\Delta u_1+\Delta v_3u_1^{(0)}
                       +\Delta v_3\Delta u_1}{t^2}
 +(1-v_3^{(0)}u_1^{(0)})
          \left(\frac1{t^2}-\frac1{t_{\rm core}^2}\right).\label{eq:exact-fluctuation-t}
\end{align}
They are equivalent to \eqref{eq:F-adjoints-c0}-\eqref{eq:F-t-c0}, with the source $\delta^2c_iI_N$ added to the adjoint equations.

\subsubsection*{Reduction and remaining F-equations}

In the block gauge \eqref{eq:exact-block-gauge}, the off-diagonal blocks of $A_2,A_3$ enter $W$ through a bilinear term from $-\Tr A_1[A_2,A_3]$ and linear terms from $-v_2A_2u_2-v_3A_3u_3$. At a critical point of $B(A_2,A_3)+L_2(A_2)+L_3(A_3)$, with $B$ bilinear and $L_2$, $L_3$ linear, one has $B=-L_2=-L_3$, so the sum equals $L_2=L_3$. Here $L_2=-\sum_s\bigl[(v_2^{\rm core}+\Delta v_2^{\rm c})\Delta A_2^{{\rm c}s}\Delta u_2^s+\Delta v_2^s\Delta A_2^{s{\rm c}}(u_2^{\rm core}+\Delta u_2^{\rm c})\bigr]-\sum_{\ell\ne m}\Delta v_2^\ell\Delta A_2^{\ell m}\Delta u_2^m$. Using $L_3$ instead gives the equivalent form of $W_{\rm links}$ with $\Delta A_3^{\ell m}$. Substituting \eqref{eq:exact-core-connecting-series} gives
$$
 \sum_s\Bigl[-\Delta v_2^sG(A_1^{ss})\Delta u_3^s
              +\Delta v_3^sH(A_1^{ss})\Delta u_2^s\Bigr]
 +W_{\rm links},
$$
with $G,H$ defined in \eqref{eq:exact-core-resolvent-functions}. Adding the direct couplings $\sum_{i,j}\Delta v_i^sK_{ij}(t)\Delta u_j^s-\sum_i\Delta v_i^sA_i^{ss}\Delta u_i^s$ gives $\Delta\mathbf v_sM_s(A^{ss})\Delta\mathbf u_s$ in the reduced superpotential \eqref{eq:exact-reduced-superpotential}. Differentiating $W_{\rm red}$ with respect to any retained field gives the corresponding original F-equation evaluated on the eliminated solution of off diagonal adjoints.

The block gauge removes the off-diagonal blocks of $A_1$, so the off-diagonal blocks of the first equation in \eqref{eq:exact-fluctuation-adj} do not follow from varying $W_{\rm red}$. They hold automatically. Invariance of $W$ under $\delta A_i=[h,A_i]$, $\delta u_i=hu_i$, $\delta v_i=-v_ih$ implies that, once the second and third adjoint equations and all column and row equations hold,
$$
 \bigl[A_1,\,[A_2,A_3]+u_1v_1+\delta^2c_1I_N\bigr]=0.
$$
Since $A_1$ is block diagonal with disjoint diagonal spectra, the off-diagonal blocks of $[A_2,A_3]+u_1v_1+\delta^2c_1I_N$ vanish.

The remaining F-equations follow from varying $W_{\rm red}$. Varying $A_2^{ss}$ and $A_3^{ss}$ gives \eqref{eq:reduced-exact-CM}, while varying $A_1^{ss}$ gives
\begin{equation}\label{eq:exact-reduced-A1}
\begin{aligned}
0={}&[A_2^{ss},A_3^{ss}]+\Delta u_1^s\Delta v_1^s+\delta^2c_1I_{n_s}
 +\Delta A_2^{s\mathrm c}\Delta A_3^{\mathrm cs}-\Delta A_3^{s\mathrm c}\Delta A_2^{\mathrm cs}\\
 &+\sum_{m\ne s}\bigl(\Delta A_2^{sm}\Delta A_3^{ms}-\Delta A_3^{sm}\Delta A_2^{ms}\bigr).
\end{aligned}
\end{equation}
The core-block products arise from varying $G(A_1^{ss})$ and $H(A_1^{ss})$ in $W_{\rm red}$, and the block-block products from varying $W_{\rm links}$. In both terms, the off-diagonal adjoints use \eqref{eq:exact-core-connecting-series} and \eqref{eq:DeltaA2DeltaA3}. Varying the rows and columns of block $s$ gives
\begin{equation}\label{eq:exact-reduced-framing}
 M_s(A^{ss})\Delta\mathbf u_s
       +\frac{\partial W_{\rm links}}{\partial\Delta\mathbf v_s}=0,
 \qquad
 \Delta\mathbf v_sM_s(A^{ss})
       +\frac{\partial W_{\rm links}}{\partial\Delta\mathbf u_s}=0.
\end{equation}
Since $W_{\rm links}$ is quartic in the strings, it enters \eqref{eq:exact-reduced-framing} starting from order $\delta^3$. The core and $t$ equations are the derivatives of $W_{\rm red}$ with respect to the core fluctuations.

\subsection{Order $\delta$ and $\delta^2$ equations}
\label{app:second-order-blocks}
\subsubsection*{Core block equations}
Varying $\mathcal Q_{\rm core}$ in \eqref{app:zero:core-quadratic} gives the order $\delta$ core equations
\begin{subequations}\label{eq:first-core-system}
\begin{align}
0={}&[A_j^{\rm core},A_k^{\mathrm{cc},(1)}]
    +[A_j^{\mathrm{cc},(1)},A_k^{\rm core}]
    +u_i^{\rm core}v_i^{\mathrm c,(1)}
    +u_i^{\mathrm c,(1)}v_i^{\rm core},\quad (i,j,k)\ \text{cyclic},
\\
0={}&A_i^{\rm core}u_i^{\mathrm c,(1)}
    +A_i^{\mathrm{cc},(1)}u_i^{\rm core}
    -\sum_j K_{ij}u_j^{\mathrm c,(1)}
    -t^{(1)}\sum_jK'_{ij}u_j^{\rm core},
\\
0={}&v_i^{\mathrm c,(1)}A_i^{\rm core}
    +v_i^{\rm core}A_i^{\mathrm{cc},(1)}
    -\sum_j v_j^{\mathrm c,(1)}K_{ji}
    -t^{(1)}\sum_jv_j^{\rm core}K'_{ji},
\\
0={}&v_1^{\mathrm c,(1)}u_3^{\rm core}
    +v_1^{\rm core}u_3^{\mathrm c,(1)}
    -\frac{v_3^{\mathrm c,(1)}u_1^{\rm core}
            +v_3^{\rm core}u_1^{\mathrm c,(1)}}{t_{\rm core}^2}
\notag\\[-1mm]
 &+\frac{2(v_3^{\rm core}u_1^{\rm core}-1)}{t_{\rm core}^3}
       t^{(1)}.
\end{align}
\end{subequations}
where $K=K(t_{\rm core})$ and $K'=\partial_tK(t_{\rm core})$, whose only nonzero entries are $K'_{13}=1$ and $K'_{31}=-t_{\rm core}^{-2}$. By Appendix~\ref{app:core-mass}, the linear operator in \eqref{eq:first-core-system} is invertible once the gauge directions are removed.

At order $\delta^2$, the same linear operator acts on the second-order core coefficients, with the inhomogeneous terms added to left hand side of the adjoint, column, row and $t$ equations, respectively
\begin{equation}\label{eq:second-core-sources}
\begin{aligned}
 \text{adjoint:}\quad&c_iI_R+\sum_s\left(
 A_j^{\mathrm cs,(1)}A_k^{s\mathrm c,(1)}
 -A_k^{\mathrm cs,(1)}A_j^{s\mathrm c,(1)}\right),\\
 \text{column:}\quad&\sum_s A_i^{\mathrm cs,(1)}u_i^{s,(1)},\\
 \text{row:}\quad&\sum_s v_i^{s,(1)}A_i^{s\mathrm c,(1)},\\
 t:\quad&\sum_s\left(v_1^{s,(1)}u_3^{s,(1)}
          -t_{\rm core}^{-2}v_3^{s,(1)}u_1^{s,(1)}\right),
\end{aligned}
\end{equation}
Here the first-order core-block adjoints follow from \eqref{eq:exact-core-connecting-series} and \eqref{eq:core-leading-order},
\begin{equation}\label{eq:first-connecting-response}
\begin{aligned}
 A_2^{\mathrm cs,(1)}
   &=-(A_1^{\rm core}-a_{1,s}I_R)^{-1}u_3^{\rm core}v_3^{s,(1)},\\
 A_3^{\mathrm cs,(1)}
   &=(A_1^{\rm core}-a_{1,s}I_R)^{-1}u_2^{\rm core}v_2^{s,(1)},\\
 A_2^{s\mathrm c,(1)}
   &=u_3^{s,(1)}v_3^{\rm core}(A_1^{\rm core}-a_{1,s}I_R)^{-1},\\
 A_3^{s\mathrm c,(1)}
   &=-u_2^{s,(1)}v_2^{\rm core}(A_1^{\rm core}-a_{1,s}I_R)^{-1}.
\end{aligned}
\end{equation}
Therefore, the order $\delta^2$ core equations fix the second-order core responses after fixing gauge directions.

\subsubsection*{The mass matrix and its kernels}
The core enters $M_s(a_{i,s})$ only through $G_s$ and $H_s$. With the resolvent convention \eqref{eq:core-resolvents}, $G_s=-v_3^{\rm core}R_1(a_{1,s})u_2^{\rm core}$ and $H_s=-v_2^{\rm core}R_1(a_{1,s})u_3^{\rm core}$, so \eqref{eq:core-six-functions} gives
\[
 p_1(a_{1,s})=a_{1,s}(1+H_s)-t_{\rm core}^{-1},\qquad
 q_1(a_{1,s})=a_{1,s}(1+G_s)-t_{\rm core}.
\]
Expanding the determinant of \eqref{eq:first-uv-quadratic},
\[
 \det M_s(a_{i,s})=-a_{1,s}a_{2,s}a_{3,s}+a_{2,s}-a_{3,s}
 -\bigl[a_{1,s}(1+G_s)(1+H_s)-t_{\rm core}(1+H_s)-t_{\rm core}^{-1}(1+G_s)\bigr].
\]
For $a_{1,s}\neq0$, the identity $p_1q_1=z^2+1$ of \eqref{eq:core-pq-differential} at $z=a_{1,s}$ sets the bracket to $a_{1,s}$. We therefore get \eqref{eq:second-scalar-surface}, which is indepedent of the core. For the empty core, $G_s=H_s=0$ and $t_{\rm core}+t_{\rm core}^{-1}=0$ give the same result.

On the surface $\det M_s(a_{i,s})=0$ the gradient of the determinant does not vanish. By Jacobi's formula below, the adjugate of $M_s(a_{i,s})$ is then nonzero on the surface, so $M_s(a_{i,s})$ has rank two. The right and left kernel vectors of $M_s(a_{i,s})$ are
\begin{equation}\label{eq:second-kernel-components}
\begin{aligned}
 \mathbf r_s&=(r_{1,s},r_{2,s},1)^{\mathsf T},&
 r_{1,s}&=\frac{a_{2,s}t_{\rm core}+1+G_s}{a_{1,s}a_{2,s}+1},&
 r_{2,s}&=\frac{t_{\rm core}-a_{1,s}(1+G_s)}{a_{1,s}a_{2,s}+1},\\
 \boldsymbol\ell_s&=(\ell_{1,s},\ell_{2,s},1),&
 \ell_{1,s}&=\frac{a_{2,s}t_{\rm core}^{-1}+1+H_s}{a_{1,s}a_{2,s}+1},&
 \ell_{2,s}&=\frac{a_{1,s}(1+H_s)-t_{\rm core}^{-1}}{a_{1,s}a_{2,s}+1}.
\end{aligned}
\end{equation}
with the third components normalized to one.
The adjugate is proportional to $\mathbf r_s\boldsymbol\ell_s$, and comparing $(3,3)$ entries gives $\operatorname{adj}M_s(a_{i,s})=(a_{1,s}a_{2,s}+1)\,\mathbf r_s\boldsymbol\ell_s$. Using Jacobi's formula $\partial\det M=\Tr(\operatorname{adj}M\,\partial M)$, we get
\begin{equation}\label{eq:resolvent-jacobi}
 \boldsymbol\ell_s\frac{\partial M_s}{\partial a_{k,s}}\mathbf r_s
 =\frac{1}{a_{1,s}a_{2,s}+1}\frac{\partial\det M_s}{\partial a_{k,s}}.
\end{equation}
The vectors in \eqref{eq:second-kernel-components} solve the first two rows of $M_s(a_{i,s})\mathbf r_s=0$ and the first two columns of $\boldsymbol\ell_sM_s(a_{i,s})=0$ also off the surface, provided $a_{1,s}a_{2,s}+1\neq0$. Eliminating the first two columns and rows leaves the Schur complement of the upper-left $2\times2$ block,
\begin{equation}\label{eq:second-schur-bilinear}
 \left.\mathbf v_s^{(1)}M_s(a_{i,s})\mathbf u_s^{(1)}\right|_{\text{first two pairs eliminated}}
   =v_3^{s,(1)}\frac{\det M_s(a_{i,s})}{a_{1,s}a_{2,s}+1}u_3^{s,(1)}.
\end{equation}
This explains the mass term for light strings in \eqref{eq:reduced-site-leading-W}.

\subsubsection*{Order $\delta^2$ block equations}
At order $\delta^2$, the leading terms of \eqref{eq:exact-reduced-A1}, \eqref{eq:reduced-exact-CM} and \eqref{eq:exact-reduced-framing} are
\begin{subequations}\label{eq:order2fullequations}
\begin{align}
0={}&[A_2^{ss,(1)},A_3^{ss,(1)}]+u_1^{s,(1)}v_1^{s,(1)}\notag\\
 &+G'_su_3^{s,(1)}v_2^{s,(1)}
 -H'_su_2^{s,(1)}v_3^{s,(1)}+c_1I_{n_s},\\
0={}&[A_3^{ss,(1)},A_1^{ss,(1)}]
 +u_2^{s,(1)}v_2^{s,(1)}+c_2I_{n_s},\\
0={}&[A_1^{ss,(1)},A_2^{ss,(1)}]
 +u_3^{s,(1)}v_3^{s,(1)}+c_3I_{n_s},\\
\sum_j(M_s)_{ij}u_j^{s,(2)}
 ={}&-\sum_{j,k}\frac{\partial(M_s)_{ij}}{\partial a_{k,s}}
                  A_k^{ss,(1)}u_j^{s,(1)},\\
\sum_jv_j^{s,(2)}(M_s)_{ji}
 ={}&-\sum_{j,k}v_j^{s,(1)}A_k^{ss,(1)}
                  \frac{\partial(M_s)_{ji}}{\partial a_{k,s}},
\end{align}
\end{subequations}
where
\begin{equation}
\begin{aligned}
G'_s=\frac{\partial G_s}{\partial a_{1,s}}
 &=v_3^{\rm core}(A_1^{\rm core}-a_{1,s}I_R)^{-2}u_2^{\rm core},\\
H'_s=\frac{\partial H_s}{\partial a_{1,s}}
 &=v_2^{\rm core}(A_1^{\rm core}-a_{1,s}I_R)^{-2}u_3^{\rm core}.
\end{aligned}
\end{equation}
The $G'_s$ and $H'_s$ terms are the core-block products in \eqref{eq:exact-reduced-A1}, evaluated on \eqref{eq:first-connecting-response}. In the last two lines, the $a_{1,s}$-derivatives also act on $G_s$ and $H_s$. Substituting $\mathbf u_s^{(1)}=\mathbf r_su_3^{s,(1)}$ and $\mathbf v_s^{(1)}=v_3^{s,(1)}\boldsymbol\ell_s$ into the three adjoint equations and using \eqref{eq:resolvent-jacobi} and \eqref{eq:second-position-conditions} gives the first three lines of \eqref{eq:second-light-system}. For example, the coefficient of $u_3^{s,(1)}v_3^{s,(1)}$ in the first line is $r_{1,s}\ell_{1,s}+G'_s\ell_{2,s}-H'_sr_{2,s}=-\boldsymbol\ell_s(\partial M_s/\partial a_{1,s})\mathbf r_s=(a_{2,s}a_{3,s}+1)/(a_{1,s}a_{2,s}+1)=c_1/c_3$, and in the second line it is $r_{2,s}\ell_{2,s}=(a_{1,s}a_{3,s}-1)/(a_{1,s}a_{2,s}+1)=c_2/c_3$. Since $M_s(a_{i,s})$ has rank two, the column and row equations are solvable precisely when their right hand sides are orthogonal to the kernels $\boldsymbol\ell_s$ and $\mathbf r_s$, respectively. By \eqref{eq:resolvent-jacobi}, contracting them with the kernels gives $\beta_s^{(1)}u_3^{s,(1)}/(a_{1,s}a_{2,s}+1)=0$ and $v_3^{s,(1)}\beta_s^{(1)}/(a_{1,s}a_{2,s}+1)=0$, which is the last line of \eqref{eq:second-light-system}. When they hold, inverting the upper-left $2\times2$ block of $M_s(a_{i,s})$ gives the massive second-order components
\begin{equation}\label{eq:second-uv-response}
\begin{aligned}
u_1^{s,(2)}={}&r_{1,s}u_3^{s,(2)}
 +\frac{-a_{2,s}A_1^{ss,(1)}u_1^{s,(1)}
         +A_2^{ss,(1)}u_2^{s,(1)}
         +G'_sA_1^{ss,(1)}u_3^{s,(1)}}{a_{1,s}a_{2,s}+1},\\
u_2^{s,(2)}={}&r_{2,s}u_3^{s,(2)}
 -\frac{A_1^{ss,(1)}u_1^{s,(1)}
         +a_{1,s}A_2^{ss,(1)}u_2^{s,(1)}
         +a_{1,s}G'_sA_1^{ss,(1)}u_3^{s,(1)}}{a_{1,s}a_{2,s}+1},\\
v_1^{s,(2)}={}&\ell_{1,s}v_3^{s,(2)}
 +\frac{-a_{2,s}v_1^{s,(1)}A_1^{ss,(1)}
         -v_2^{s,(1)}A_2^{ss,(1)}
         +H'_sv_3^{s,(1)}A_1^{ss,(1)}}{a_{1,s}a_{2,s}+1},\\
v_2^{s,(2)}={}&\ell_{2,s}v_3^{s,(2)}
 +\frac{v_1^{s,(1)}A_1^{ss,(1)}
         -a_{1,s}v_2^{s,(1)}A_2^{ss,(1)}
         +a_{1,s}H'_sv_3^{s,(1)}A_1^{ss,(1)}}{a_{1,s}a_{2,s}+1}.
\end{aligned}
\end{equation}
The third column and row equations then hold by the last line of \eqref{eq:second-light-system}. The kernel components $u_3^{s,(2)}$ and $v_3^{s,(2)}$ remain free at this order.

\subsection{Order $\delta^3$ equations}
\label{app:quartic-reduction}

This appendix derives \eqref{eq:third-schur-expansion}, and the coefficient $e_s$ and $\mathcal T_s$ in $\mathcal W_{\rm eff}^{(4)}$ \eqref{eq:third-effective-W}.

For the column-row coupling with matrix-valued adjoints, eliminating the first two column-row pairs of $\Delta\mathbf v_sM_s(A^{ss})\Delta\mathbf u_s$ leaves
\begin{equation}
 \Delta v_3^s\Bigl[(M_s)_{33}-\sum_{a,b=1}^2(M_s)_{3a}\bigl[(M_s)_{\le2}^{-1}\bigr]_{ab}(M_s)_{b3}\Bigr]\Delta u_3^s,
\end{equation}
where every entry is evaluated on $A^{ss}$ with the core at its background value, and $(M_s)_{\le2}$ is the upper-left $2\times2$ block. At second order, this elimination is the substitution of \eqref{eq:second-uv-response}. The coefficients of order $\delta^0$ and $\delta$ of the bracket vanish by $\det M_s(a_{i,s})=0$ and $\beta_s^{(1)}=0$, so the second-order kernel components $u_3^{s,(2)}$ and $v_3^{s,(2)}$ drop out of $\mathcal W^{(4)}$. For commuting arguments, the bracket equals $\det M_s/(a_{1,s}a_{2,s}+1)$ as in \eqref{eq:second-schur-bilinear}, and its order $\delta^2$ part gives the first two terms of \eqref{eq:third-schur-expansion}. For matrices, solving the first two rows gives the bracket in ordered form,
$$
 -A_3^{ss}+t\bigl(I+H(A_1^{ss})\bigr)
 +\bigl[t^{-1}-\bigl(I+H(A_1^{ss})\bigr)A_1^{ss}\bigr]
 \bigl(I+A_2^{ss}A_1^{ss}\bigr)^{-1}
 \bigl(I+G(A_1^{ss})+tA_2^{ss}\bigr),
$$
with $t=t_{\rm core}$. Its order $\delta^2$ term contains products of two light adjoints in a definite order, from $(I+A_2^{ss}A_1^{ss})^{-1}$ and the products of $H(A_1^{ss})A_1^{ss}, G(A_1^{ss})$ with $A_2^{ss}$. Since $A_1^{ss,(1)}=-(c_2A_2^{ss,(1)}+c_3A_3^{ss,(1)})/c_1$, each product differs from its symmetrization by a multiple of $[A_2^{ss,(1)},A_3^{ss,(1)}]$. Collecting these multiples gives the last term of \eqref{eq:third-schur-expansion}, with
\[
 \frac{c_1}{c_3}e_s=\frac{c_3}{2c_1(a_{1,s}a_{2,s}+1)^2}\bigl(p_1'q_1-p_1q_1'\bigr)\big|_{z=a_{1,s}},
\]
where $p_1, q_1$ are given in Appendix~\ref{app:second-order-blocks}. By \eqref{eq:core-pq-differential}, $p_1'q_1-p_1q_1'=p_1q_1(p_1'/p_1-q_1'/q_1)=2\kappa$. With \eqref{eq:second-position-conditions},
\[
 e_s=\frac{c_3^2\,\kappa}{c_1^2(a_{1,s}a_{2,s}+1)^2}=\frac{\kappa}{(a_{2,s}a_{3,s}+1)^2}.
\]

The other contributions to $\mathcal W^{(4)}$ come from the core and from $W_{\rm links}$. By \eqref{eq:second-core-sources}, \eqref{eq:first-connecting-response} and \eqref{eq:second-kernel-components}, the second-order core responses are linear in the $c_i$ and in the contractions $v_3^{s,(1)}u_3^{s,(1)}$. At leading order, $W_{\rm links}$ is a sum of products of the same contractions. Together with the commutator term of \eqref{eq:third-schur-expansion}, these contributions take the form
\begin{equation}
\begin{aligned}
 \mathcal W_{\rm corr}^{(4)}={}&
 \sum_s e_s\frac{c_1}{c_3}v_3^{s,(1)}
       [A_2^{ss,(1)},A_3^{ss,(1)}]u_3^{s,(1)}\\
 &+\frac{c_1^2}{c_3^2}\sum_{s,s'}b_{ss'}
       \bigl(v_3^{s,(1)}u_3^{s,(1)}\bigr)
       \bigl(v_3^{s',(1)}u_3^{s',(1)}\bigr)\\
 &+\frac{c_1}{c_3}\sum_s d_sv_3^{s,(1)}u_3^{s,(1)}+\text{constant}.
\end{aligned}
\end{equation}
where the coefficients $b_{ss'}=b_{s's}$ and $d_s$ are fixed by the core and block-block responses and are not needed. Using $\frac{c_1}{c_3}v_3^{s,(1)}u_3^{s,(1)}=\Tr\Theta_s-c_1n_s$ and
\[
 \frac{c_1}{c_3}v_3^{s,(1)}[A_2^{ss,(1)},A_3^{ss,(1)}]u_3^{s,(1)}
 =\Tr\!\left(\frac{c_1}{c_3}u_3^{s,(1)}v_3^{s,(1)}\Theta_s\right)
 -\frac{c_1^2}{c_3^2}\bigl(v_3^{s,(1)}u_3^{s,(1)}\bigr)^2
 -\frac{c_1^2}{c_3}v_3^{s,(1)}u_3^{s,(1)},
\]
we find
\begin{equation}\label{eq:third-multiplier-shift}
\begin{aligned}
 \mathcal W_{\rm corr}^{(4)}
   &=\sum_s\Tr(\mathcal T_s\Theta_s)+\mathrm{const},\\
 \mathcal T_s={}&e_s\frac{c_1}{c_3}u_3^{s,(1)}v_3^{s,(1)}
       +\Bigl[d_s+c_1e_s(2n_s-1)
                 -2c_1\sum_{s'} b_{ss'}n_{s'}\\
 &\hspace{25mm}
       +\sum_{s'}b_{ss'}\Tr\Theta_{s'}-e_s\Tr\Theta_s\Bigr]I_{n_s}.
\end{aligned}
\end{equation}
These terms shift the Lagrange multiplier in \eqref{eq:third-effective-W} by $\mathcal T_s$. On $\Theta_s=0$ they reduce to a constant fixed by the occupation numbers, so they do not affect the counting.

\section{Nondegeneracy of the core quadratic form}
\label{app:core-mass}
In this appendix we show that every solution of the order $\delta$ core equations \eqref{eq:first-core-system} is an infinitesimal $\GL(R,\C)$ transformation with $t^{(1)}=0$. Hence $\mathcal Q_{\rm core}$ in \eqref{app:zero:core-quadratic} is nondegenerate once the gauge directions are removed and all physical core fluctuations are massive. Consider a nonempty core as constructed in Section~\ref{sec:ZeroSourceSolutions}. The empty core is simple and treated below \eqref{eq:core-leading-order}. Suppressing the core labels, we write $A_i^{(1)}$, $u_i^{(1)}$, $v_i^{(1)}$, $t^{(1)}$ for an arbitrary solution of \eqref{eq:first-core-system}. Any function of the core fields is expanded similarly with the spectral parameter $z$ held fixed.

Since the order $\delta$ core equations \eqref{eq:first-core-system} do not contain source terms or scalar-block fields, they are precisely the linearization of the zero-source core F-equations. Consequently, the identities \eqref{eq:core-pq-differential} for $p_i,q_i$ remain valid through order $\delta$.
Expanding $p=p_1$ and $\kappa$ near the background value gives
$$
 p(z;\delta)=p(z)+\delta p^{(1)}(z)+O(\delta^2),\qquad
 \kappa(\delta)=\kappa+\delta\kappa^{(1)}+O(\delta^2).
$$
Using definition \eqref{eq:core-kappa}, we find $\kappa^{(1)}=v_2^{(1)}u_3+v_2u_3^{(1)}-\frac{t^{(1)}}{t^2}$. Substituting the expansions into $p'/p=(z+\kappa)/(z^2+1)$, at order $\delta$ we find
$$
 \frac{d}{dz}\left(\frac{p^{(1)}}{p}\right)
 =\frac{\kappa^{(1)}}{z^2+1},
$$
Both $p$ and $p^{(1)}$ are rational functions of $z$, since they are obtained from the finite-dimensional matrix resolvent. The derivative of the rational function $p^{(1)}/p$ has zero residue at every finite pole. On the other hand, the right-hand side has residues $\kappa^{(1)}/(2\i)$ and $-\kappa^{(1)}/(2\i)$ at $z=\i$ and $z=-\i$. Hence $\kappa^{(1)}=0$ and $p^{(1)}/p$ is constant. To fix this constant, use the large-$z$ expansion
$$
 R_1(z;\delta)=\frac{I_R}{z}+\frac{A_1(\delta)}{z^2}+O(z^{-3}).
$$
The definition of $p_1$ then gives
$$
 \begin{aligned}
 p(z;\delta)&=z-t(\delta)^{-1}-v_2(\delta)u_3(\delta)+O(z^{-1})\\
             &=z-\kappa(\delta)+O(z^{-1}),\\
 p^{(1)}(z)&=-\kappa^{(1)}+O(z^{-1}).
 \end{aligned}
$$
Since $p^{(1)}$ is bounded as $z\to\infty$ with $\kappa^{(1)}=0$ and $p\sim z$, the constant $p^{(1)}/p$ must vanish, hence $p^{(1)}=0$.
Varying $p_1$ in \eqref{eq:core-six-functions} with $p^{(1)}=0$ gives
$$
 \bigl(v_2R_1(z)u_3\bigr)^{(1)}=\frac{t^{(1)}}{t^2z}.
$$
Since $A_1$ has eigenvalues $\pm\i$ by \eqref{eq:core-jordan-form-derived}, it is invertible. Its resolvent $R_1(z)$ and its first-order variation $R_1(z)A_1^{(1)}R_1(z)$ are therefore regular at $z=0$, so the left hand side is regular. The right-hand side can then have no pole, which forces $t^{(1)}=0$.

We now apply the reconstruction argument of Section~\ref{sec:ZeroSourceSolutions} to first order. Since $\kappa^{(1)}=0$, linearizing \eqref{eq:core-pq-differential} for all three adjoints gives
$$
 \frac{d}{dz}\left(\frac{p_i^{(1)}}{p_i}\right)=0,
 \qquad
 \frac{d}{dz}\left(\frac{q_i^{(1)}}{q_i}\right)=0.
$$
By the same large-$z$ argument as above, $p_i^{(1)}=q_i^{(1)}=0$ for every $i$. With $t^{(1)}=0$, expanding the definitions \eqref{eq:core-six-functions} at large $z$ then gives
$$
 \bigl(v_iA_k^nu_j\bigr)^{(1)}=0,
 \qquad n\ge0,\quad i,j,k\ \text{distinct}.
$$
The contraction argument near \eqref{eq:contraction} also holds to first order. Roughly, every contraction $v_i P(A_1,A_2,A_3)u_j$ is reduced to products of the contractions $v_iA_k^nu_j$ with $i,j,k$ distinct, with coefficients that depend only on $t$. Linearizing this with $t^{(1)}=0$ gives
\begin{equation}
 \bigl(v_iP(A_1,A_2,A_3)u_j\bigr)^{(1)}=0,
 \label{eq:core-contractions-first-order}
\end{equation}
for every polynomial $P$.

By the core construction, the columns $P(A)u_i$ span $\C^R$, and the rows $v_jQ(A)$ span its dual. Define a linear map $h$ on the spanning columns by
\[
 h\bigl(P(A)u_i\bigr)=\bigl(P(A)u_i\bigr)^{(1)}.
\]
To check that $h$ is well-defined, consider any relation $s=\sum_i P_i(A)u_i=0$, with fixed polynomials $P_i$. By \eqref{eq:core-contractions-first-order} and the product rule,
\[
 0=\bigl(v_jQ(A)s\bigr)^{(1)}=v_jQ(A)s^{(1)},
\]
where the term $\bigl(v_jQ(A)\bigr)^{(1)}s$ vanishes because $s=0$. Since the rows $v_jQ(A)$ span the dual space, $s^{(1)}=0$. Thus every linear relation among the spanning columns is preserved to first order, and $h$ is well-defined.

For $w=P(A)u_i$, the product rule and \eqref{eq:core-contractions-first-order} now give
\[
 \begin{aligned}
 h(A_kw)&=(A_kw)^{(1)}=A_k^{(1)}w+A_khw,\\
 0&=(v_jw)^{(1)}=v_j^{(1)}w+v_jhw.
 \end{aligned}
\]
Because these columns $w$ span the core, the first identity gives $A_k^{(1)}=[h,A_k]$ and the second gives $v_j^{(1)}=-v_jh$. Taking $P=1$ also gives $u_i^{(1)}=hu_i$. Every solution of \eqref{eq:first-core-system} is therefore
\[
 A_i^{(1)}=[h,A_i],\qquad u_i^{(1)}=hu_i,
 \qquad v_i^{(1)}=-v_i h,\qquad t^{(1)}=0.
\]
Distinct $h$ generate distinct solutions, since if $[h,A_i]=0$ and $hu_i=0$, $h$ annihilates every $P(A)u_i$ and vanishes. The kernel of $\mathcal Q_{\rm core}$ then has dimension $R^2$ out of the $3R^2+6R+1$ core coordinates, and $\mathcal Q_{\rm core}$ is nondegenerate on the remaining $2R^2+6R+1$ directions once $\GL(R,\C)$ is fixed.

\section{Explicit partition solutions}
\label{app:partition-solutions}

This appendix collects explicit solutions of modified Calogero-Moser space \eqref{eq:third-modified-CM} labeled by partitions $\nu_s$. Besides the multiplier spectrum \eqref{eq:third-partition-spectrum}, Propositions~6 and~8 of \cite{Felder:2024thr} give
\begin{equation}\label{eq:partition-QL-spectrum}
\begin{aligned}
 \operatorname{Spec}\mathsf Q_s
   &=\{\text{zeros of }\mathcal H_{\nu_s}(z)\},\qquad
 \operatorname{Spec}\mathsf L_s=\i\,\operatorname{Spec}\mathsf Q_s,\\
 \mathcal H_{\nu_s}(z)
   &=\operatorname{Wr}\!\left(H_{\nu_{s,1}+\ell-1},\ldots,H_{\nu_{s,\ell}}\right)(z).
\end{aligned}
\end{equation}
Here $\operatorname{Wr}(f_1,\ldots,f_\ell)=\det(\partial_z^{a-1}f_b)_{a,b=1}^{\ell}$ is the Wronskian and $H_m(z)=(-1)^me^{z^2}\partial_z^me^{-z^2}$ is the Hermite polynomial, so that $\mathcal H_{\nu_s}$ has degree $n_s$. When the zeros $x_1,\ldots,x_{n_s}$ of $\mathcal H_{\nu_s}$ are distinct, Proposition~4 of \cite{Felder:2024thr} gives the solution explicitly. Suppressing the label $s$, a representative of the gauge orbit is
\begin{equation}\label{eq:third-partition-profiles}
\begin{gathered}
 \mathsf Q=\operatorname{diag}(x_1,\ldots,x_{n_s}),\qquad
 \mathsf v=(1,\ldots,1)^T,\qquad \mathsf w=(1,\ldots,1),\\
 \mathsf L_{ab}=\begin{cases}
  (x_a-x_b)^{-1},&a\ne b,\\0,&a=b,
 \end{cases}
 \qquad
 \mathsf M_{ab}=\begin{cases}
  -(x_a-x_b)^{-2},&a\ne b,\\
  \displaystyle\sum_{c\ne a}(x_a-x_c)^{-2},&a=b,
 \end{cases}
\end{gathered}
\end{equation}
When zeros coincide, $\mathsf Q$ cannot be diagonalized. Wilson's construction reviewed in \cite[Section~3]{Felder:2024thr} supplies representatives in this case.

As an example, the following matrices represent the three partitions of $n_s=3$, with $e_a$ the standard basis columns:
\begin{equation}\label{eq:third-partition-three}
\begin{gathered}
 \mathsf L=\begin{pmatrix}0&\tfrac12&0\\\rho&0&\tfrac12\\0&\sigma&0\end{pmatrix},
 \qquad
 \mathsf Q=\begin{pmatrix}0&-\tfrac12&0\\\rho&0&-\tfrac12\\0&\sigma&0\end{pmatrix},\\[3pt]
 \begin{array}{c|rr|c|c|c}
 \nu_s&\rho&\sigma&\mathsf M&\mathsf v&\mathsf w\\ \hline
 (3)&-2&-1&\operatorname{diag}(0,1,2)&3e_1&e_1^T\\
 (2,1)&1&-1&\operatorname{diag}(-1,0,1)&3e_2&e_2^T\\
 (1,1,1)&1&2&\operatorname{diag}(-2,-1,0)&3e_3&e_3^T
 \end{array}
\end{gathered}
\end{equation}
One checks directly that they satisfy \eqref{eq:third-modified-CM}. The spectra of $\mathsf Q$ are $\{0,\pm\sqrt{3/2}\}$, $\{0,0,0\}$ and $\{0,\pm\i\sqrt{3/2}\}$. They are the zeros of $\mathcal H_{(3)}\propto z(2z^2-3)$, $\mathcal H_{(2,1)}\propto z^3$ and $\mathcal H_{(1,1,1)}\propto z(2z^2+3)$. The solution is nondegenerate even though the eigenvalues coincide.  The physical leading fields follow by inverting \eqref{eq:third-paper-map}, with $u_3^{s,(1)}=-c_3\mathsf v_s$ and $v_3^{s,(1)}=\mathsf w_s$. The remaining first-order fields follow from $A_1^{ss,(1)}=-(c_2A_2^{ss,(1)}+c_3A_3^{ss,(1)})/c_1$, the kernel relations \eqref{eq:second-kernel-components} and the core-block adjoints \eqref{eq:first-connecting-response}. $\Lambda_s=c_3\omega_s\mathsf M_s$ fixes the second-order normal combination in \eqref{eq:third-effective-W}.

\end{appendix}

\addcontentsline{toc}{section}{References}
\bibliographystyle{JHEP}
\bibliography{bibliography}

\providecommand{\href}[2]{#2}\begingroup\raggedright\begin{thebibliography}{10}

\bibitem{Preskill:1991tb}
J.~Preskill, P.~Schwarz, A.~D. Shapere, S.~Trivedi and F.~Wilczek, \emph{Limitations on the statistical description of black holes}, \href{https://doi.org/10.1142/S0217732391002773}{\emph{Mod. Phys. Lett. A} {\bfseries 6} (1991) 2353}.

\bibitem{Iliesiu:2020qvm}
L.~V. Iliesiu and G.~J. Turiaci, \emph{The statistical mechanics of near-extremal black holes}, \href{https://doi.org/10.1007/JHEP05(2021)145}{\emph{JHEP} {\bfseries 05} (2021) 145} [\href{https://arxiv.org/abs/2003.02860}{{\ttfamily 2003.02860}}].

\bibitem{Heydeman:2020hhw}
M.~Heydeman, L.~V. Iliesiu, G.~J. Turiaci and W.~Zhao, \emph{The statistical mechanics of near-bps black holes}, \href{https://doi.org/10.1088/1751-8121/ac3be9}{\emph{J. Phys. A} {\bfseries 55} (2022) 014004} [\href{https://arxiv.org/abs/2011.01953}{{\ttfamily 2011.01953}}].

\bibitem{Cabo-Bizet:2018ehj}
A.~Cabo-Bizet, D.~Cassani, D.~Martelli and S.~Murthy, \emph{Microscopic origin of the bekenstein-hawking entropy of supersymmetric {AdS}$_5$ black holes}, \href{https://doi.org/10.1007/JHEP10(2019)062}{\emph{JHEP} {\bfseries 10} (2019) 062} [\href{https://arxiv.org/abs/1810.11442}{{\ttfamily 1810.11442}}].

\bibitem{Iliesiu:2021are}
L.~V. Iliesiu, M.~Kologlu and G.~J. Turiaci, \emph{Supersymmetric indices factorize}, \href{https://doi.org/10.1007/JHEP05(2023)032}{\emph{JHEP} {\bfseries 05} (2023) 032} [\href{https://arxiv.org/abs/2107.09062}{{\ttfamily 2107.09062}}].

\bibitem{Lin:2022zxd}
H.~W. Lin, J.~Maldacena, L.~Rozenberg and J.~Shan, \emph{Looking at supersymmetric black holes for a very long time}, \href{https://doi.org/10.21468/SciPostPhys.14.5.128}{\emph{SciPost Phys.} {\bfseries 14} (2023) 128} [\href{https://arxiv.org/abs/2207.00408}{{\ttfamily 2207.00408}}].

\bibitem{Chen:2024oqv}
Y.~Chen, H.~W. Lin and S.~H. Shenker, \emph{{BPS} chaos}, \href{https://doi.org/10.21468/SciPostPhys.18.2.072}{\emph{SciPost Phys.} {\bfseries 18} (2025) 072} [\href{https://arxiv.org/abs/2407.19387}{{\ttfamily 2407.19387}}].

\bibitem{Chang:2024zqi}
C.-M. Chang and Y.-H. Lin, \emph{Holographic covering and the fortuity of black holes},  \href{https://arxiv.org/abs/2402.10129}{{\ttfamily 2402.10129}}.

\bibitem{Boruch:2023gfn}
J.~Boruch, L.~V. Iliesiu, S.~Murthy and G.~J. Turiaci, \emph{New forms of attraction: attractor saddles for the black hole index}, \href{https://doi.org/10.1007/JHEP04(2025)087}{\emph{JHEP} {\bfseries 04} (2025) 087} [\href{https://arxiv.org/abs/2310.07763}{{\ttfamily 2310.07763}}].

\bibitem{Chen:2024gmc}
Y.~Chen, S.~Murthy and G.~J. Turiaci, \emph{Gravitational index of the heterotic string}, \href{https://doi.org/10.1007/JHEP09(2024)041}{\emph{JHEP} {\bfseries 09} (2024) 041} [\href{https://arxiv.org/abs/2402.03297}{{\ttfamily 2402.03297}}].

\bibitem{Boruch:2025biv}
J.~Boruch, L.~V. Iliesiu, S.~Murthy and G.~J. Turiaci, \emph{Multicentered black hole saddles for supersymmetric indices},  \href{https://arxiv.org/abs/2507.07166}{{\ttfamily 2507.07166}}.

\bibitem{Gaiotto:2005rp}
D.~Gaiotto, M.~Guica, L.~Huang, A.~Simons, A.~Strominger and X.~Yin, \emph{{D4-D0 branes on the quintic}},  \href{https://arxiv.org/abs/hep-th/0509168}{{\ttfamily hep-th/0509168}}.

\bibitem{Gaiotto:2006wm}
D.~Gaiotto, A.~Strominger and X.~Yin, \emph{{The M5-Brane Elliptic Genus: Modularity and BPS States}}, \href{https://doi.org/10.1088/1126-6708/2007/08/070}{\emph{JHEP} {\bfseries 08} (2007) 070} [\href{https://arxiv.org/abs/hep-th/0607010}{{\ttfamily hep-th/0607010}}].

\bibitem{Denef:2007vg}
F.~Denef and G.~W. Moore, \emph{{Split states, entropy enigmas, holes and halos}}, \href{https://doi.org/10.1007/JHEP11(2011)129}{\emph{JHEP} {\bfseries 11} (2011) 129} [\href{https://arxiv.org/abs/hep-th/0702146}{{\ttfamily hep-th/0702146}}].

\bibitem{Alexandrov:2022pgd}
S.~Alexandrov, N.~Gaddam, J.~Manschot and B.~Pioline, \emph{{Modular bootstrap for D4-D2-D0 indices on compact Calabi{\textendash}Yau threefolds}}, \href{https://doi.org/10.4310/ATMP.2023.v27.n3.a2}{\emph{Adv. Theor. Math. Phys.} {\bfseries 27} (2023) 683} [\href{https://arxiv.org/abs/2204.02207}{{\ttfamily 2204.02207}}].

\bibitem{Alexandrov:2023ltz}
S.~Alexandrov, S.~Feyzbakhsh, A.~Klemm and B.~Pioline, \emph{{Quantum geometry and mock modularity}}, \href{https://doi.org/10.4310/cntp.260401215658}{\emph{Commun. Num. Theor. Phys.} {\bfseries 20} (2026) 97} [\href{https://arxiv.org/abs/2312.12629}{{\ttfamily 2312.12629}}].

\bibitem{Alexandrov:2023zjb}
S.~Alexandrov, S.~Feyzbakhsh, A.~Klemm, B.~Pioline and T.~Schimannek, \emph{{Quantum geometry, stability and modularity}}, \href{https://doi.org/10.4310/CNTP.2024.v18.n1.a2}{\emph{Commun. Num. Theor. Phys.} {\bfseries 18} (2024) 49} [\href{https://arxiv.org/abs/2301.08066}{{\ttfamily 2301.08066}}].

\bibitem{Maldacena:1999bp}
J.~M. Maldacena, G.~W. Moore and A.~Strominger, \emph{{Counting BPS black holes in toroidal Type II string theory}},  \href{https://arxiv.org/abs/hep-th/9903163}{{\ttfamily hep-th/9903163}}.

\bibitem{Shih:2005qf}
D.~Shih, A.~Strominger and X.~Yin, \emph{{Counting dyons in N=8 string theory}}, \href{https://doi.org/10.1088/1126-6708/2006/06/037}{\emph{JHEP} {\bfseries 06} (2006) 037} [\href{https://arxiv.org/abs/hep-th/0506151}{{\ttfamily hep-th/0506151}}].

\bibitem{Pioline:2005vi}
B.~Pioline, \emph{{BPS black hole degeneracies and minimal automorphic representations}}, \href{https://doi.org/10.1088/1126-6708/2005/08/071}{\emph{JHEP} {\bfseries 08} (2005) 071} [\href{https://arxiv.org/abs/hep-th/0506228}{{\ttfamily hep-th/0506228}}].

\bibitem{Chowdhury:2014yca}
A.~Chowdhury, R.~S. Garavuso, S.~Mondal and A.~Sen, \emph{{BPS State Counting in N=8 Supersymmetric String Theory for Pure D-brane Configurations}}, \href{https://doi.org/10.1007/JHEP10(2014)186}{\emph{JHEP} {\bfseries 10} (2014) 186} [\href{https://arxiv.org/abs/1405.0412}{{\ttfamily 1405.0412}}].

\bibitem{Chowdhury:2015gbk}
A.~Chowdhury, R.~S. Garavuso, S.~Mondal and A.~Sen, \emph{{Do All BPS Black Hole Microstates Carry Zero Angular Momentum?}}, \href{https://doi.org/10.1007/JHEP04(2016)082}{\emph{JHEP} {\bfseries 04} (2016) 082} [\href{https://arxiv.org/abs/1511.06978}{{\ttfamily 1511.06978}}].

\bibitem{Chowdhury:2023wss}
A.~Chowdhury and S.~Maji, \emph{{Counting $\mathcal{N}$ = 8 black holes as algebraic varieties}}, \href{https://doi.org/10.1007/JHEP05(2024)091}{\emph{JHEP} {\bfseries 05} (2024) 091} [\href{https://arxiv.org/abs/2311.04786}{{\ttfamily 2311.04786}}].

\bibitem{Chowdhury:2026noq}
A.~Chowdhury and S.~Maji, \emph{{Pure D-brane black holes: BPS counting and non-BPS vacua}}, \href{https://doi.org/10.1007/JHEP08(2026)070}{\emph{JHEP} {\bfseries 08} (2026) 070} [\href{https://arxiv.org/abs/2601.07927}{{\ttfamily 2601.07927}}].

\bibitem{Denef:2002ru}
F.~Denef, \emph{{Quantum quivers and Hall / hole halos}}, \href{https://doi.org/10.1088/1126-6708/2002/10/023}{\emph{JHEP} {\bfseries 10} (2002) 023} [\href{https://arxiv.org/abs/hep-th/0206072}{{\ttfamily hep-th/0206072}}].

\bibitem{Hori:2014tda}
K.~Hori, H.~Kim and P.~Yi, \emph{{Witten Index and Wall Crossing}}, \href{https://doi.org/10.1007/JHEP01(2015)124}{\emph{JHEP} {\bfseries 01} (2015) 124} [\href{https://arxiv.org/abs/1407.2567}{{\ttfamily 1407.2567}}].

\bibitem{Sen:2009gy}
A.~Sen, \emph{{Arithmetic of N=8 Black Holes}}, \href{https://doi.org/10.1007/JHEP02(2010)090}{\emph{JHEP} {\bfseries 02} (2010) 090} [\href{https://arxiv.org/abs/0908.0039}{{\ttfamily 0908.0039}}].

\bibitem{King94}
A.~D. King, \emph{Moduli of representations of finite dimensional algebras}, \href{https://doi.org/10.1093/qmath/45.4.515}{\emph{The Quarterly Journal of Mathematics} {\bfseries 45} (1994) 515}.

\bibitem{Hoskins2014}
V.~Hoskins, \emph{Stratifications associated to reductive group actions on affine spaces}, \href{https://doi.org/10.1093/qmath/hat046}{\emph{The Quarterly Journal of Mathematics} {\bfseries 65} (2014) 1011} [\href{https://arxiv.org/abs/1210.6811}{{\ttfamily 1210.6811}}].

\bibitem{CB01}
W.~Crawley-Boevey, \emph{Geometry of the moment map for representations of quivers}, \href{https://doi.org/10.1023/A:1017558904030}{\emph{Compositio Mathematica} {\bfseries 126} (2001) 257}.

\bibitem{Przezdziecki:2016}
T.~Przezdziecki, \emph{The combinatorics of {$\mathbb{C}^*$}-fixed points in generalized {Calogero--Moser} spaces and {Hilbert} schemes}, \href{https://doi.org/10.1016/j.jalgebra.2020.04.003}{\emph{Journal of Algebra} {\bfseries 556} (2020) 936} [\href{https://arxiv.org/abs/1610.03920}{{\ttfamily 1610.03920}}].

\bibitem{Wilson}
G.~Wilson, \emph{Collisions of {Calogero--Moser} particles and an adelic {Grassmannian}}, \href{https://doi.org/10.1007/s002220050237}{\emph{Inventiones Mathematicae} {\bfseries 133} (1998) 1} With an appendix by I. G. Macdonald.

\bibitem{Felder:2024thr}
G.~Felder, A.~P. Veselov and N.~Nekrasov, \emph{{Harmonic Locus and Calogero-Moser Spaces}}, \href{https://doi.org/10.1007/s00220-025-05522-z}{\emph{Commun. Math. Phys.} {\bfseries 407} (2026) 22} [\href{https://arxiv.org/abs/2404.18471}{{\ttfamily 2404.18471}}].

\bibitem{Sen:2008ta}
A.~Sen, \emph{{N=8 Dyon Partition Function and Walls of Marginal Stability}}, \href{https://doi.org/10.1088/1126-6708/2008/07/118}{\emph{JHEP} {\bfseries 07} (2008) 118} [\href{https://arxiv.org/abs/0803.1014}{{\ttfamily 0803.1014}}].

\bibitem{Iliesiu:2022kny}
L.~V. Iliesiu, S.~Murthy and G.~J. Turiaci, \emph{{Black hole microstate counting from the gravitational path integral}}, \href{https://doi.org/10.1007/JHEP08(2025)152}{\emph{JHEP} {\bfseries 08} (2025) 152} [\href{https://arxiv.org/abs/2209.13602}{{\ttfamily 2209.13602}}].

\bibitem{Banerjee:2011jp}
S.~Banerjee, R.~K. Gupta, I.~Mandal and A.~Sen, \emph{{Logarithmic Corrections to N=4 and N=8 Black Hole Entropy: A One Loop Test of Quantum Gravity}}, \href{https://doi.org/10.1007/JHEP11(2011)143}{\emph{JHEP} {\bfseries 11} (2011) 143} [\href{https://arxiv.org/abs/1106.0080}{{\ttfamily 1106.0080}}].

\bibitem{Anagnostopoulos:2007fw}
K.~N. Anagnostopoulos, M.~Hanada, J.~Nishimura and S.~Takeuchi, \emph{{Monte Carlo studies of supersymmetric matrix quantum mechanics with sixteen supercharges at finite temperature}}, \href{https://doi.org/10.1103/PhysRevLett.100.021601}{\emph{Phys. Rev. Lett.} {\bfseries 100} (2008) 021601} [\href{https://arxiv.org/abs/0707.4454}{{\ttfamily 0707.4454}}].

\bibitem{wip}
C.-M. Chang, S.~Duary, X.~Shi, G.~J. Turiaci and H.~Zhang 2026.

\bibitem{Turiaci:2023jfa}
G.~J. Turiaci and E.~Witten, \emph{{$ \mathcal{N} $ = 2 JT supergravity and matrix models}}, \href{https://doi.org/10.1007/JHEP12(2023)003}{\emph{JHEP} {\bfseries 12} (2023) 003} [\href{https://arxiv.org/abs/2305.19438}{{\ttfamily 2305.19438}}].

\bibitem{Biggs:2023mfn}
A.~Biggs and J.~Maldacena, \emph{{Scaling similarities and quasinormal modes of D0 black hole solutions}}, \href{https://doi.org/10.1007/JHEP11(2023)155}{\emph{JHEP} {\bfseries 11} (2023) 155} [\href{https://arxiv.org/abs/2303.09974}{{\ttfamily 2303.09974}}].

\bibitem{Benjamin:2019stq}
N.~Benjamin, H.~Ooguri, S.-H. Shao and Y.~Wang, \emph{{Light-cone modular bootstrap and pure gravity}}, \href{https://doi.org/10.1103/PhysRevD.100.066029}{\emph{Phys. Rev. D} {\bfseries 100} (2019) 066029} [\href{https://arxiv.org/abs/1906.04184}{{\ttfamily 1906.04184}}].

\bibitem{Ghosh:2019rcj}
A.~Ghosh, H.~Maxfield and G.~J. Turiaci, \emph{{A universal Schwarzian sector in two-dimensional conformal field theories}}, \href{https://doi.org/10.1007/JHEP05(2020)104}{\emph{JHEP} {\bfseries 05} (2020) 104} [\href{https://arxiv.org/abs/1912.07654}{{\ttfamily 1912.07654}}].

\end{thebibliography}\endgroup

\end{document}